\documentclass[aps,prd,twocolumn,amsmath,amssymb,floatfix]{revtex4}
\usepackage{natbib}
\usepackage{graphicx}
\usepackage{dcolumn}
\usepackage{bm}
\usepackage{multirow}
\pdfoutput=1 
\begin{document} 
\title{Comparative study of charged multiplicities and moments in $pp$ collisions at $\sqrt{s}$=7~TeV in the forward region at the LHC. }
\author{R.~Aggarwal$^a$}
\author{M.~Kaur$^b$}
\email{manjit@pu.ac.in} 
\affiliation{\\$^a$~Department of Technology, Savitribai Phule Pune University, Pune-411 007, India.\\
                 $^b$~Department of Physics, Panjab University, Chandigarh-160 014, India.\\}
\date{\today}
\begin{abstract}
Charged particle multiplicities in proton-proton collisions measured in the LHCb detector at a centre-of-mass energy of $\sqrt s$=7~TeV in different windows of pseudorapidity $\eta$, in the forward region of the vertex detector are studied by using different statistical distributions.~Three distributions are compared with the data and the moments of the distributions are calculated.~The data constituting two sets, one of minimum bias events and another of hard QCD events are analysed.~The distributions considered derive from different functional forms based on underlying interaction dynamics.~The analysis complements the multiplicity analysis done by LHCb in terms of Monte Carlo event generators.~The present analysis is from a different perspective, using statistical distributions.
\end{abstract}  
\maketitle
\section{INTRODUCTION}
The focus of the high energy experimentation has undergone a paradiem shift from fixed target experiments to collider experiments in persuit of increasing energy in the center-of-mass system (c.m.s).~The latest experiments at the large hadron collider (LHC) have lead to several new results.~The energy available for particle production in $pp$ collisions at the LHC, results in multitude of charged particles, the number of which is often the first observable measured in all experimental set ups.~The increase in numbers follows logarithmic rise in the average values with the increasing energy of collision in center of mass system.~The number of charged particles is predicted to be connected with the underlying dynamics of interactions.~Numerous theoretical, phenomenological and statistical models have been proposed to develop an understanding of the interaction dynamics.~In high energy physics the Negative Binomial Distribution which exhibits approximate Koba-Nielsen-Olesen (KNO) scaling has been used since very early times  \cite{Kob, NBD1, NBD2, NBD3, NBD4, NBD5, NBD6, NBD7, NBD8}.~First failure of KNO was reported in the analysis of $\overline{p}p$ data obtained by UA5 collaboration \cite{Aln}, followed by similar observations made by other experiments UA1 \cite{Abe,Alba}.~As a result probability versus number of charged particles could not follow the NB behaviour, due to the appearence of a shoulder structure.~This triggered the interest in modifying the negative binomial distribution.~The first suggestion was put forth by C. Fuglesang \cite{Fug} who proposed to consider the NB distribution as composed of weighted superposition of two components, soft events (events without mini-jets) and semi-hard events (events with mini-jets).~The fraction of soft events $\alpha$ is taken as a weight and multiplicity distribution of each component being NB type.~So that the P(total distribution, NB type)= $\alpha$ P(soft event distribution-NB) + (1-$\alpha$)P(semi-hard event distribution-NB), where P stands for the probability.~Such a distribution was referred to as modified-NBD.~Since then, several of the statistical distributions have been modified in the similar way and used for describing the multiplicity distributions at different energies.~Some of these are Gamma distribution \cite{Gam1,Gam2}, Tsallis distribution \cite{TS1, TS2}, the shifted Gompertz distribution \cite{SGD1, SGD2} and the Weibull distribution \cite{WB1, WB2} for the description of particle production which successfully explain the multiplicity distributions in different kinds of collisions.~The charged particle multiplicity at the LHC has been measured by CMS, ATLAS and ALICE experiments \cite{CMS,ATLAS,ALICE} mainly in the central region.~While the LHCb  experiment is the only experiment to measure it in the forward region at $\sqrt{s}$=7~TeV \cite{LHCb} and study the multiplicity distributions in different phase space slices, in comparison to predictions from several Monte Carlo event generators.

In the present analysis, the first study of multiplicity distributions in $pp$ data collected by the LHCb collaboration in the forward region of the detector in terms of three distributions namely, negative binomial (NBD), shifted Gompertz (SGD) and Weibull (WB) distributions is reported.~The forward region spanning the psudorapidity $\eta$ range between -2.0 $<\eta<$ -2.5 and 2.0 $<\eta<$ 4.5 with a further division into smaller pseudorapidity windows is studied.~The study of the forward region is particularly interesting as region is sensitive to low Bjorken-$x$ QCD and which plays an important role in multi partonic interactions (MPI) and in the understanding of interaction dynamics.~In addition to the multiplicity distributions, the other standard physical observables are the normalized moments ($C_{q}$), the normalized factorial moments ($F_{q}$) and normalized factorial cumulants ($K_{q}$).~The ratio of the cumulants to factorial moments has also been widely studies.~We give an outline of the models used in the following section.

The paper is organised as follows; we describe the essential details required for the calculations of probability distributions NBD, SGD and WB and the data used in analysis in section II.~Section III gives results obtained from the comparison of our analysis of the three distributions, analysis of moments and cumulants of moments followed by conclusions in ~Section IV .
\section{DISTRIBUTIONS AND THE PARAMETRIZATIONS}

Charged particle multiplicity can be characterised by a function $P(n|\tilde{\omega})$, which determines the probability of producing $n$ charged particles in an interaction, given a set of parameters $\tilde{\omega}$.~The function $P(n|\tilde{\omega})$ may represent a probability distribution function (PDF) predicting the distribution of number of particles according to the distribution.~The mean of this distribution gives the average number of particles produced.~We discuss PDFs of the three distributions below;

{\bf NBD}: The following probability distribution function defines the distribution known as the
negative binomial distribution in the variable $n$;

PDF:
\small
\begin{equation}
\begin{split}
P(n|<n>,k) = \binom {n+k-1}{k-1}\left(\frac{<n>/k}{1+<n>/k}\right)^{n}\\
\times(1+<n>/k)^{-k} \
\end{split}
\end{equation}
\normalsize

in the general case, the binomial coefficient is written as $k(k + 1).....(k+n-1)/n!$ when the positive parameter $k$ is not an integer.~The first parameter, $n$ determines the position, being equal to the expected average, $<n>$, and $k$ influences the shape of the distribution.

{\bf SGD}: The Shifted Gompertz Distribution has two independent random variables, one of which has an exponential distribution with a parameter $b$ and the other has a Gumbel distribution, also known as log-Weibull distribtion, with parameters $\beta$ and $b$.~We proposed to use this distribution for studying the collision data obtained from the high energy colliders, Super proton Synchrotron (SPS), Large electron positron collider (LEP), Large Hadron collider (LHC).~These data are for $\overline{p}p$, $e^{+}e^{-}$ and $pp$ collisions at various c.m.s. energies.~In the detailed studies we have shown that SGD describes the data trends very well \cite{SGD1,SGD2,SGD3}.~To fit the data the probability density function (PDF) is described by two nonnegative free parameters, $b$, the scale parameter and $\eta$ determining the shape of the distribution.~Following equations define the distribution;

PDF:
\small
\begin{equation}
P(n|b,\beta) = be^{-bn}e^{-\beta e^{-bn}}[1+\beta(1-e^{-bn})]\hspace{0.3cm} for\hspace{0.15cm} n > 0 
\end{equation}  
Mean of the distribution:
\small
\begin{equation}
(-\frac{1}{b})(E[ln(Y)]-ln(\beta)) \hspace*{0.3 cm} where \hspace*{0.2 cm} Y=\beta e^{-bn}\\
\end{equation}
\small
\begin{equation}
\begin{split}
E[ln(Y)] = [1 + \frac{1}{\beta}]\int_{0}^{\infty}e^{-Y}[ln(Y)]dY\\ - \frac{1}{\beta}\int_{0}^{\infty}Y e^{-Y}[ln(Y)]dY\
\end{split}
\end{equation}
\normalsize
where $b\geq0$ and $\beta\geq0$.

Though the validity of SGD has been tested by us recently for the charged particle multiplicity distribution in the $pp$ collision data at $\sqrt{s}\backsim7$~TeV, collected by the CMS collaboration \cite{CMS} in the central region, this is the first analysis of the data collected by the LHCb experiment \cite{LHCb}, in the forward region.

{\bf WB}:  The charged multiplicity data from variety of collision types and energies, as mentioned before have also been analysed using Weibull distribution \cite{WB1, WB2,WB3}.~Weibull distribution is also a two parameter distribution.~In its standard form, these two parameters represent scale and shape of the distribution.~The two parameter Weibull has been used during the last few years to describe the collision data from high energy experiments.~The probability distribution function can be defined as below.

PDF:
\small
\begin{multline}
P_N(N|\lambda,K)=\begin{cases}
\frac{K}{\lambda} \left (\frac{N}{\lambda} \right )^{(K-1)} exp^{-\left (\frac{N}{\lambda}\right )^{ K}} & N \geq 0 \\                                   
0 &  N < 0 \end{cases}\\ 
\end{multline} 

\normalsize
$\lambda>0$ is a scale parameter $\lambda>0$ and $K>0$ is the shape parameter.
 
Mean of the distribution function is given by:
\small
\begin{equation}
\bar{N} = \lambda \Gamma(1+1/K)
\end{equation}
\normalsize
For a multiplicity distribution, the normalised moments $C_{q}$, normalised factorial moments ($F_{q}$), normalised factorial cumulants ($K_{q}$) and ratio of the two ($H_{q}$) moments are defined as;
\small
\begin{gather}
C_{q} = \frac{\sum_{n=1}^{\infty}n^{q}P(n)}{(\sum_{n=1}^{\infty}nP(n))^q}\\
F_{q} = \frac{\sum_{n=q}^{\infty}n(n-1).......(n-q+1)P(n)}{(\sum_{n=1}^{\infty}nP(n))^q}\\
K_{q} = F_{q}-\sum_{m=1}^{q-1}\frac{(q-1)!}{m!(q-m-1)!}K_{q-m}F_{m}\\
H_{q} = K_{q}/F_{q}\
\end{gather}

\normalsize
\subsection{THE DATA USED}
Charged particle multiplicity distributions at $\sqrt {s}$=7 TeV, collected by the LHCb collaboration \cite{LHCb} using the vertex detector (VELO) have been analysed.~The vertex detector has been so designed as to provide a uniform acceptance in the forward region with additional coverage of the backward region.~Particle multiplicity is measured using only tracks reconstructed with the VELO.~Further the tracks are considerd only if their pseudorapdity lies either in the range $-2.5<\eta<-2.0$ or $2.0<\eta<4.5$.~The measurements are done in the forward range divided in to five pseudorapidity windows with size $\Delta\eta$ = 0.5.~We have analysed the distributions in each of these windows separately.~Two samples of data are available; (i) the minimum bias events which have one or more reconstructed tracks in the vertex detector and (ii) the hard QCD events with each event having atleast one track with transverse momentum $>$1 GeV/c.

\section{RESULTS}
LHCb studied the experimentally measured charged particle multiplicity distributions in different pseudorapidity windows and also in the full forward region by comparing with several event generators \cite{LHCb}.~None are able to describe fully the multiplicity distributions as a function of $\eta$.~In general, the models were found to underestimate the data.~In the present paper we study the experimental distributions from a different perspective.

The experimental charged multiplicity distributions are studied with the PDFs from the negative binomial, shifted Gompertz and Weibull distributions.~All these distributions are two parameter distributions, namely scale and shape parameters.~The PDFs are calculated by using equations (1-6) and matching with the data by carrying out minimum $\chi^{2}$ fits using ROOT6.18.

\subsection{COMPARISON OF PDFs OF DIFFERENT DISTRIBUTIONS OF MULTIPLICITIES}

Fits to the data for minimum bias events are shown in figure~1 for five pseudorapidity windows.~Figure~2 shows the similar figures for the hard QCD events.~Tables~I-II give the parameters of the fits and the corresponding $\chi^{2}/ndf$ and $p$-values for both minimum bias events and hard QCD events, for all the distributions.     

One finds that in comparison to the minimum bias data, the multiplicity distributions for hard QCD events have larger high-multiplicity tails.~In general, the NBD, SGD and WB distributions reproduce the data very well in almost all the $\eta$ values.~The hard QCD fits are far better than the corresponding minimum bias distributions.~However, all distribution fail in the forward region with pseudorapidity range 2.0$<\eta<$4.5 with $p$-values corresponding to $CL< 0.10\%$ and all the fits are statistically excluded in the two categories of sample.~But only in the case of NBD, the multiplicity distribution of hard QCD events in 2.0$<\eta<$4.5 fits very well.~In addition, the pseudorapidity range 4.0$ < \eta < $4.5 remains poorly described in SGD for this category.~Most of these observations agree with the observations made by LHCb \cite{LHCb}, but in a different study using event generators.

~It has been well established since the observations made by UA5 collaboration \cite{Aln} that the multiplicity distributions at higher collision energies show a shoulder structure in full phase space.~This feature however is not shown by any of the three distributions.~And all the fits fail in the forward region.~As proposed by A.~Giovannini et al \cite{Gio} that the observed shoulder structure can be described by using a weighted superposition of two component distributions.~One describing the soft events distribution and another
describing the semi-hard events distribution, with each distribution following the NBD.~Adopting this approach, we redefine each of the probability distributions, NBD, SGD and WB as weighted superpositions of two component distributions and fit accordingly as follows;

\small
\begin{equation}
P(n)^{X}=\alpha P(n)_{soft}^{X}+(1-\alpha)P(n)_{semi-hard}^{X}
\end{equation}
where X stands for NBD, SGD or WB distribution.
 
\normalsize
Figure~3 shows the fits of the convolution of two component distributions and we label them as 2NBD, 2SGD and 2WB.~The corresponding fit parameters are given in table~III.~It is observed that the data fit perfectly well with each of the modified distributions with minimum $\chi^{2}/ndf$ values, due to which the $p$-values in each case turn out to be very nearly 1.0.

An interesting observation is the oscillations of $H_{q}$ as a function of the rank $q$ obtained from data.~These are reproduced by the ratio of cumulants, $K_{q}/F_{q}$ calculated from the superposition of 2NBD, 2SGD or 2WB.~This leads to the fact that the second multiplicity component is connected with cumulants, each of which involves an infinite cumulative sum over all multiplicity probabilities, as shown in equation (8).~The next section describes the moment analysis.

\begin{figure}
\includegraphics[width=3.7in, height= 2.9in]{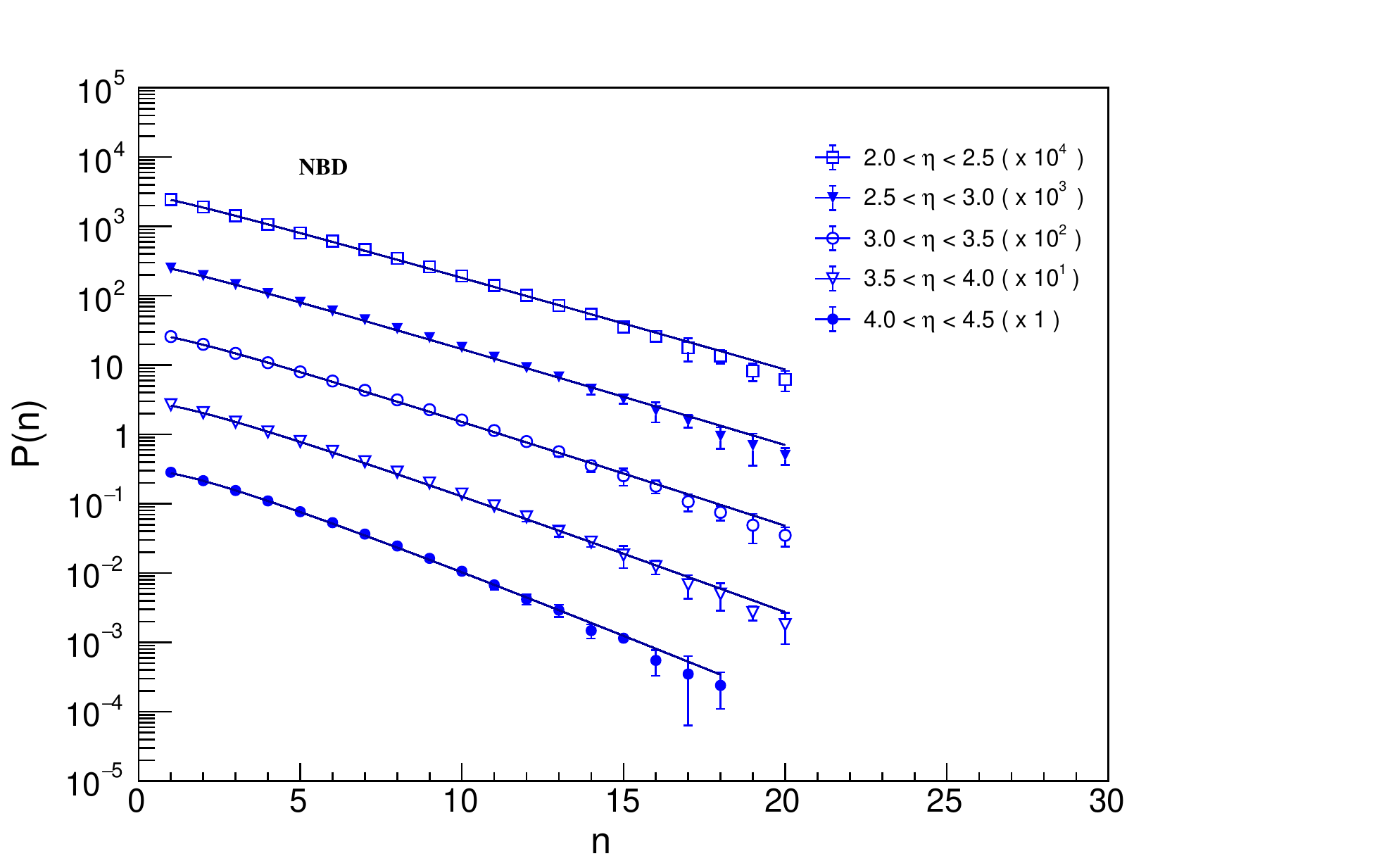}
\includegraphics[width=3.7in, height= 2.9in]{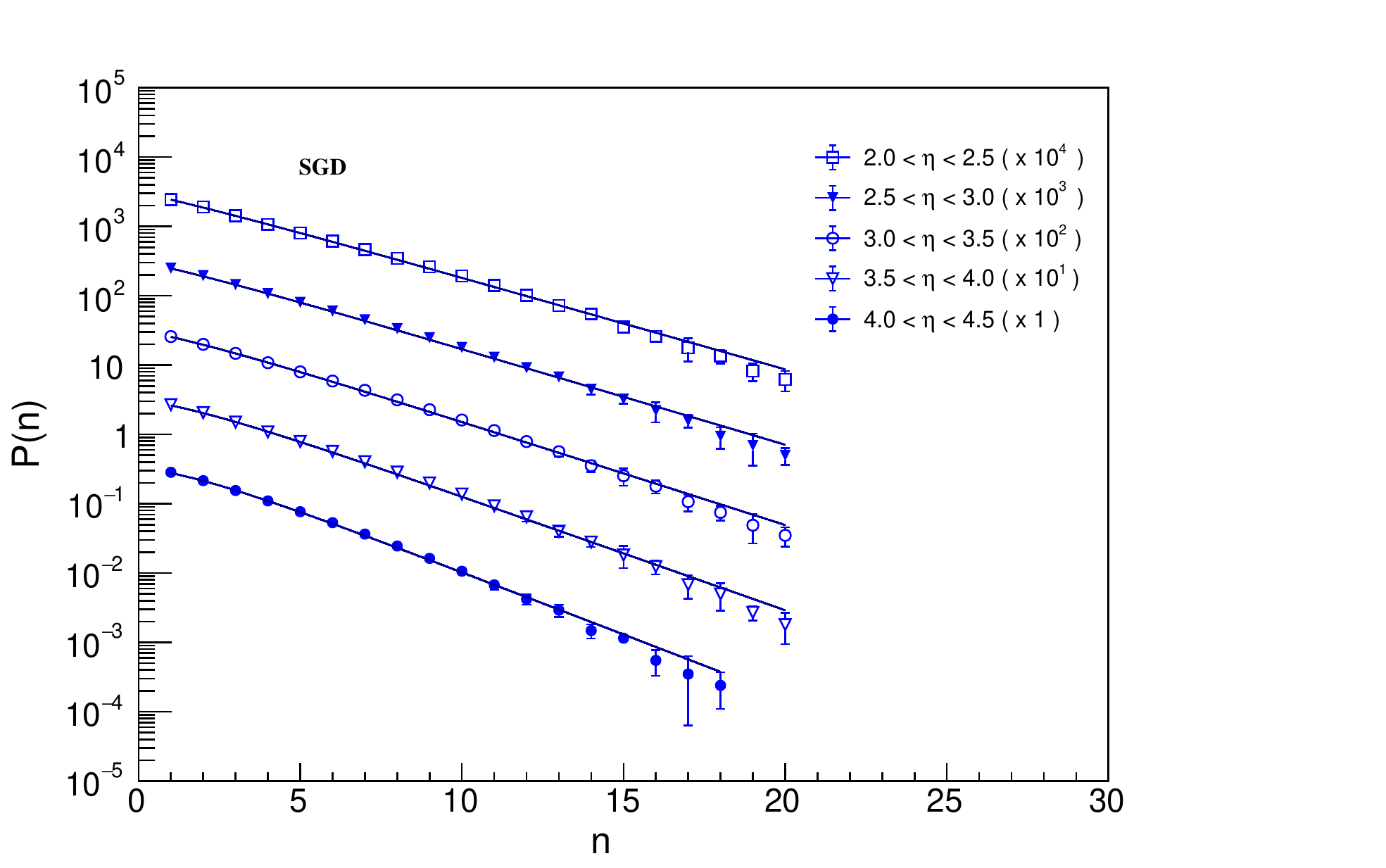}
\includegraphics[width=3.7in, height= 2.9in]{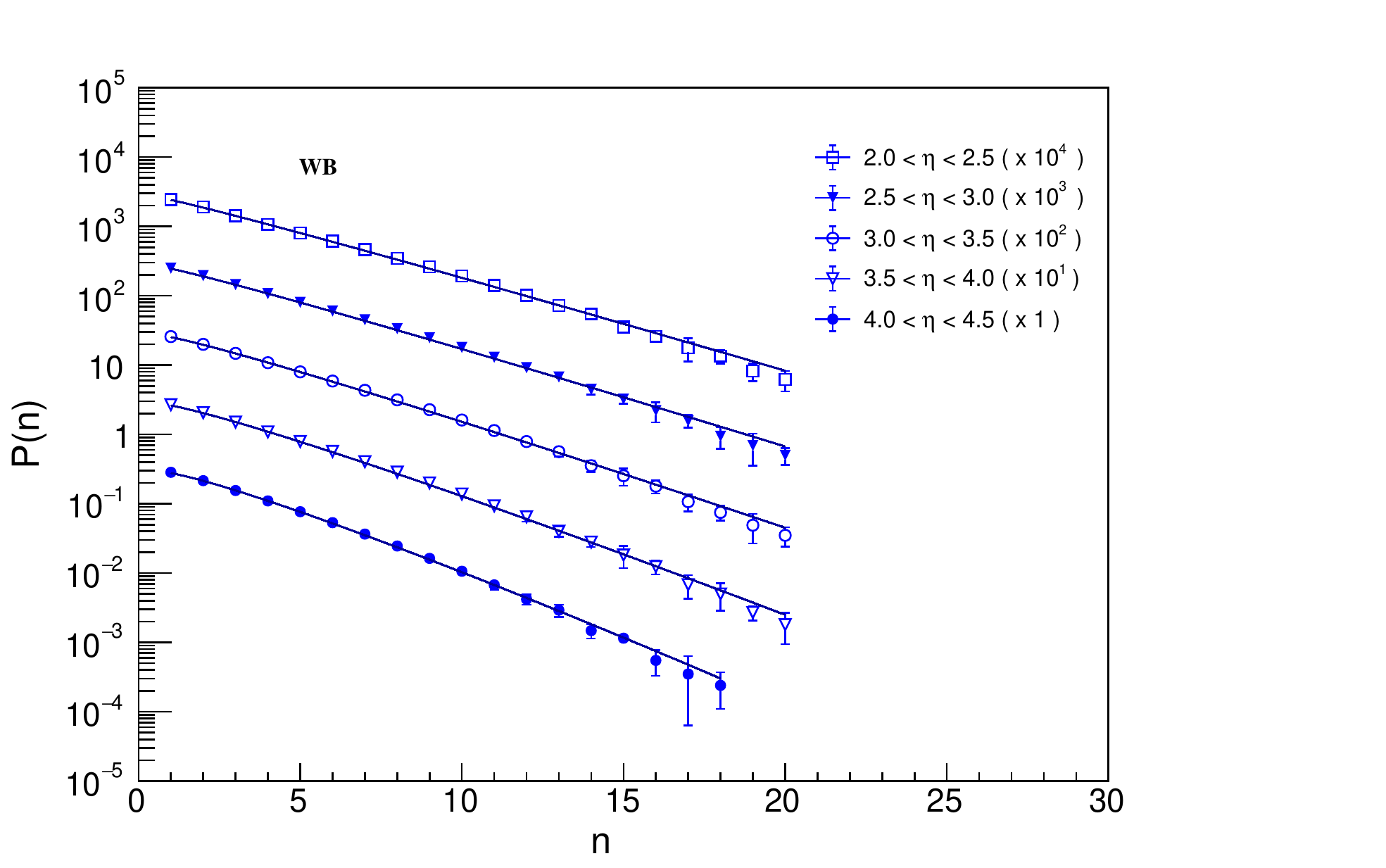}
\caption{Data on charged particle multiplicity distributions in $pp$ minimum bias events at $\sqrt{s}$= 7~TeV.~Points show the data and solid lines are the fits for various distributions (top to bottom) in different pseudorapidity intervals.}
\end{figure}

\begin{figure}
\includegraphics[width=3.7 in, height= 2.9 in]{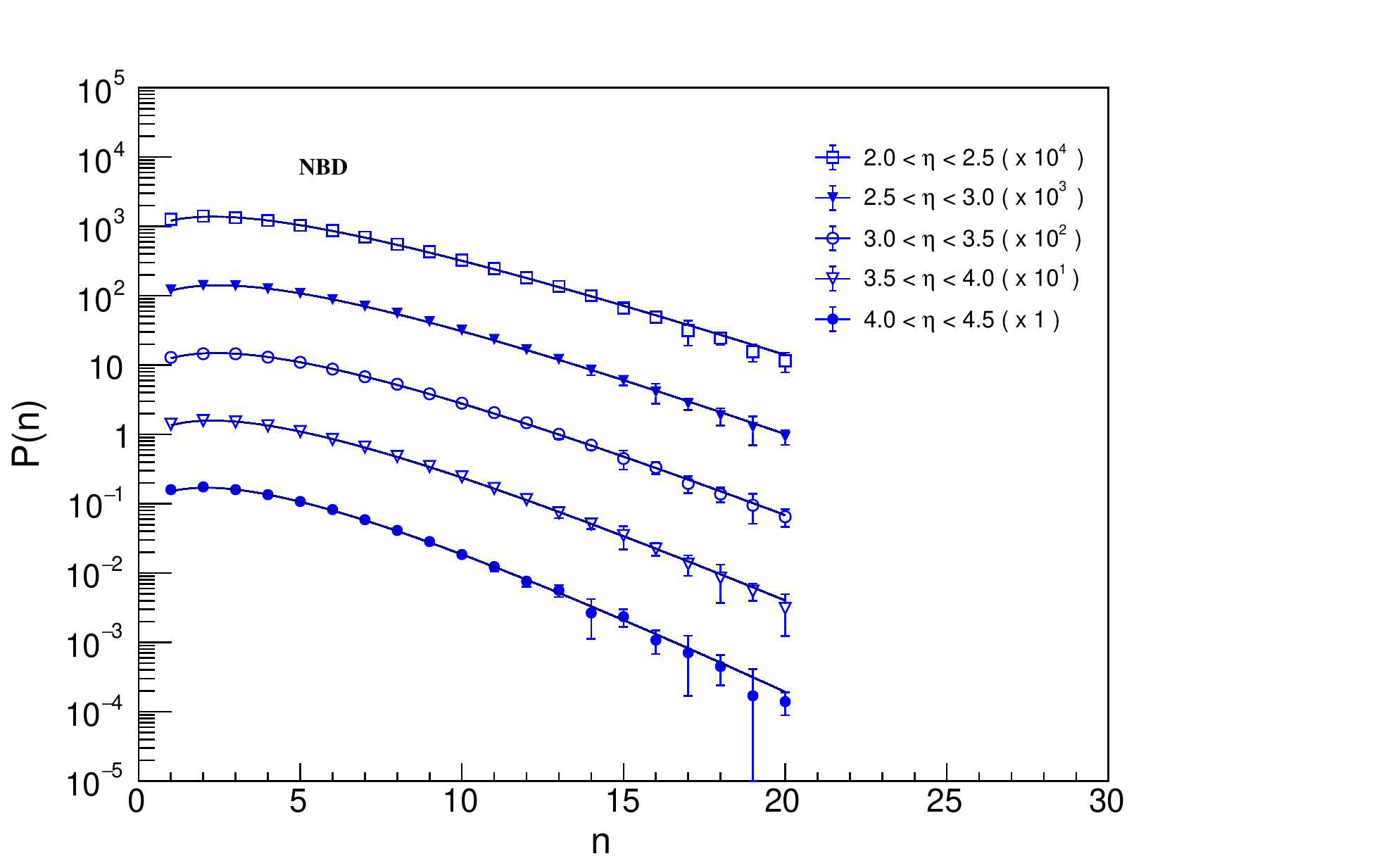}
\includegraphics[width=3.7 in, height= 2.9 in]{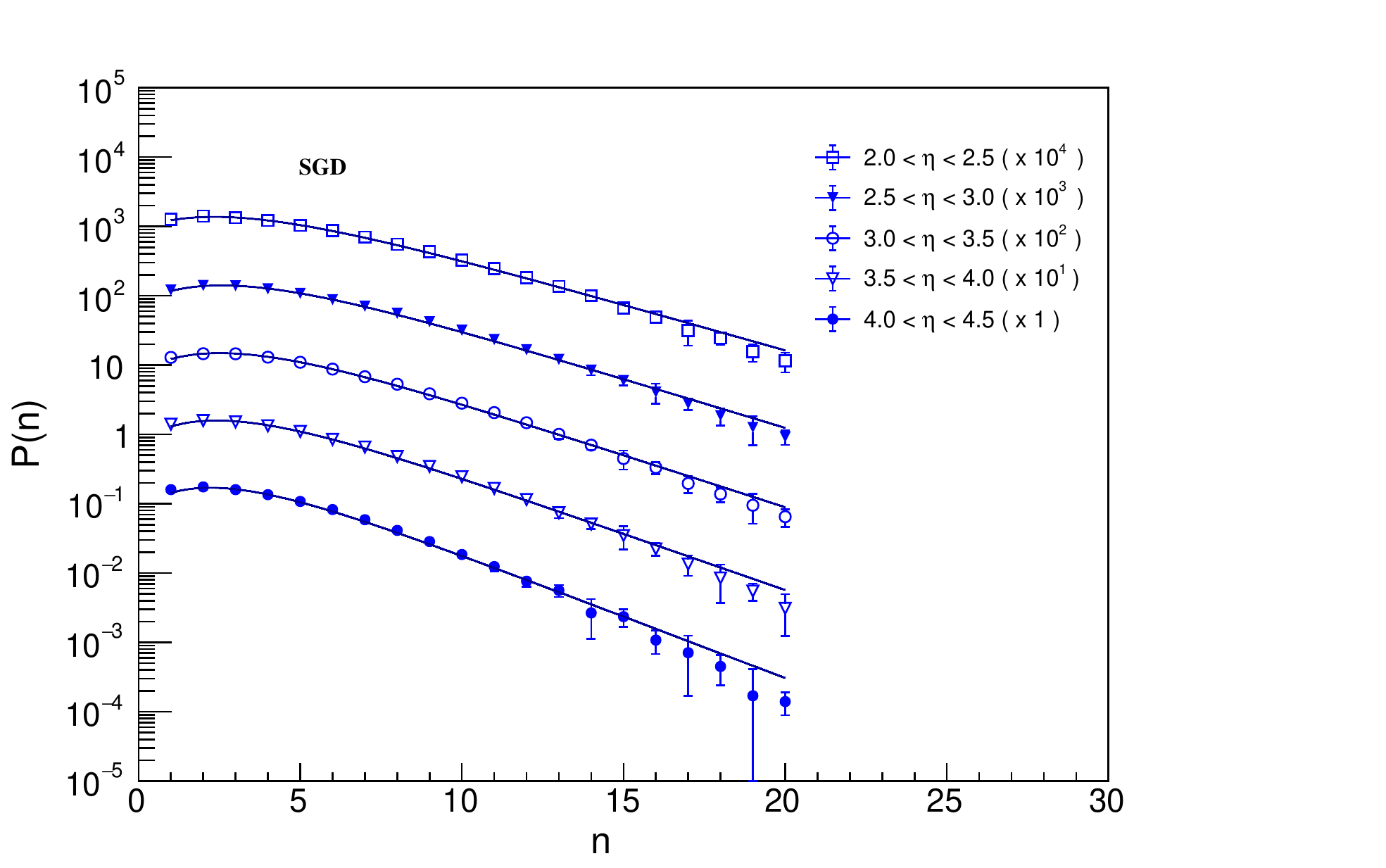}
\includegraphics[width=3.7 in, height= 2.9 in]{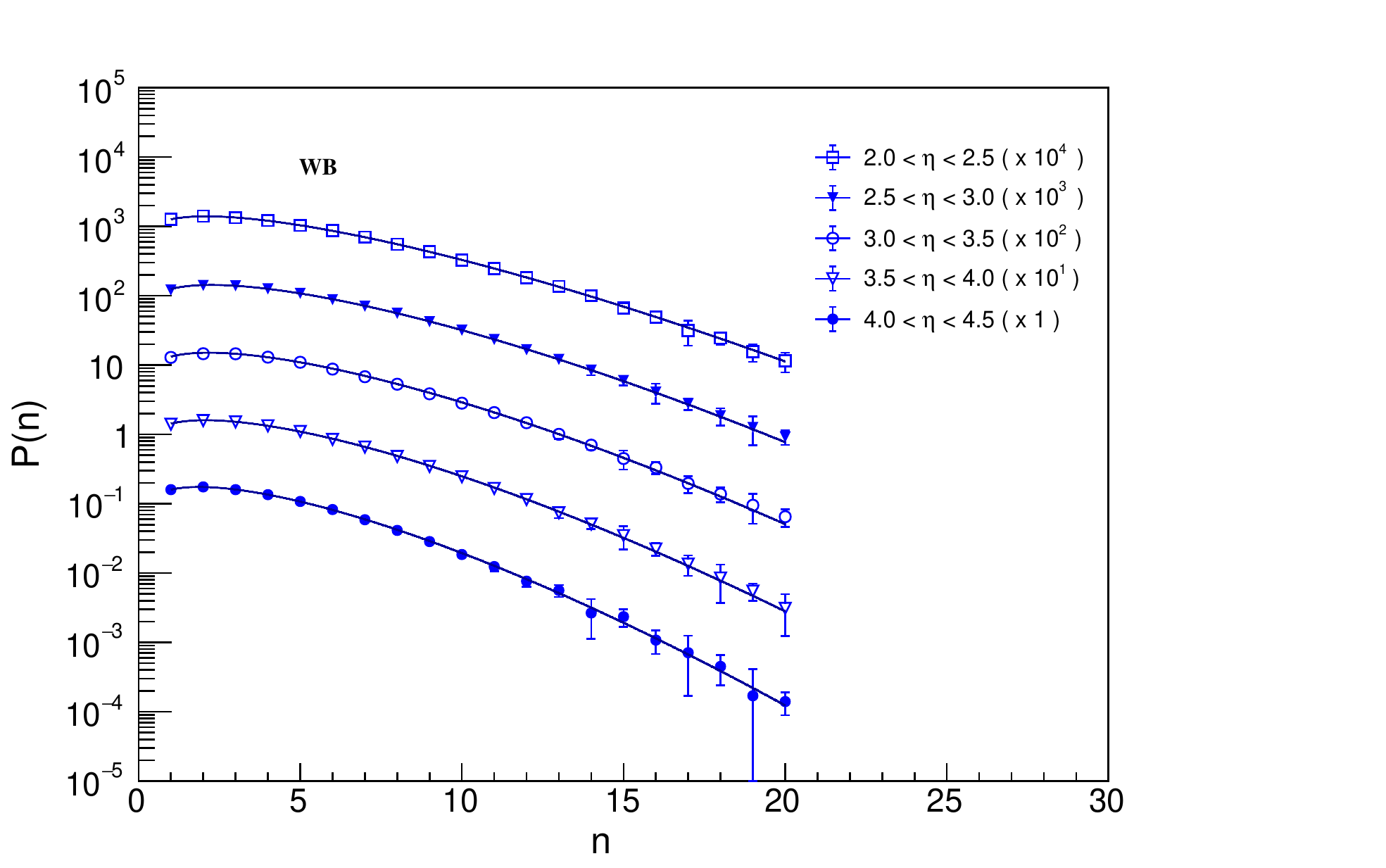}
\caption{Data on charged particle multiplicity distributions in $pp$ hard QCD events at $\sqrt{s}$= 7~TeV.~Points show the data and solid lines are the fits for various distributions (top to bottom) in different pseudorapidity intervals.}
\end{figure}

\begin{figure}
\includegraphics[width=3.7 in, height= 2.9 in]{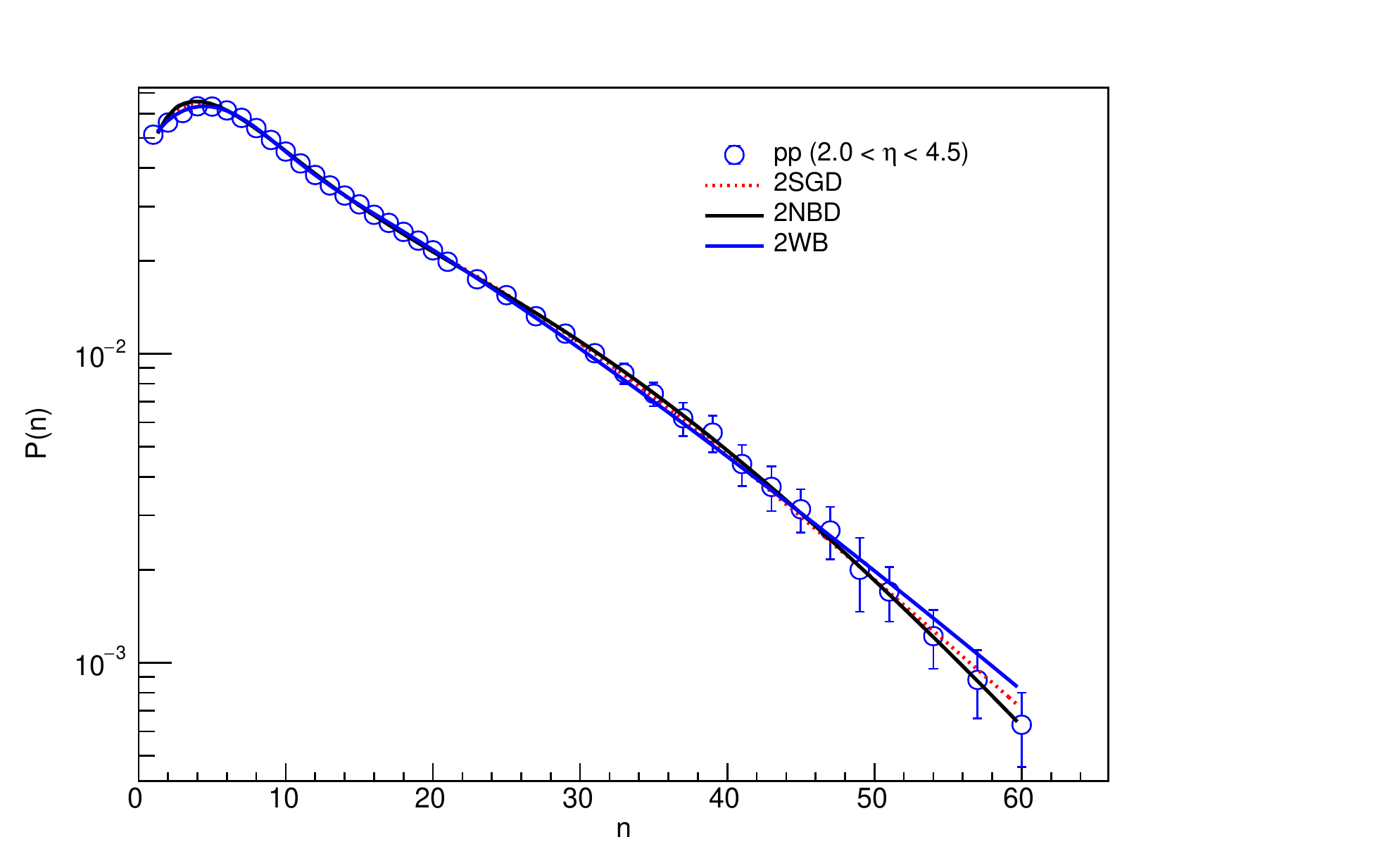}
\includegraphics[width=3.7 in, height= 2.9 in]{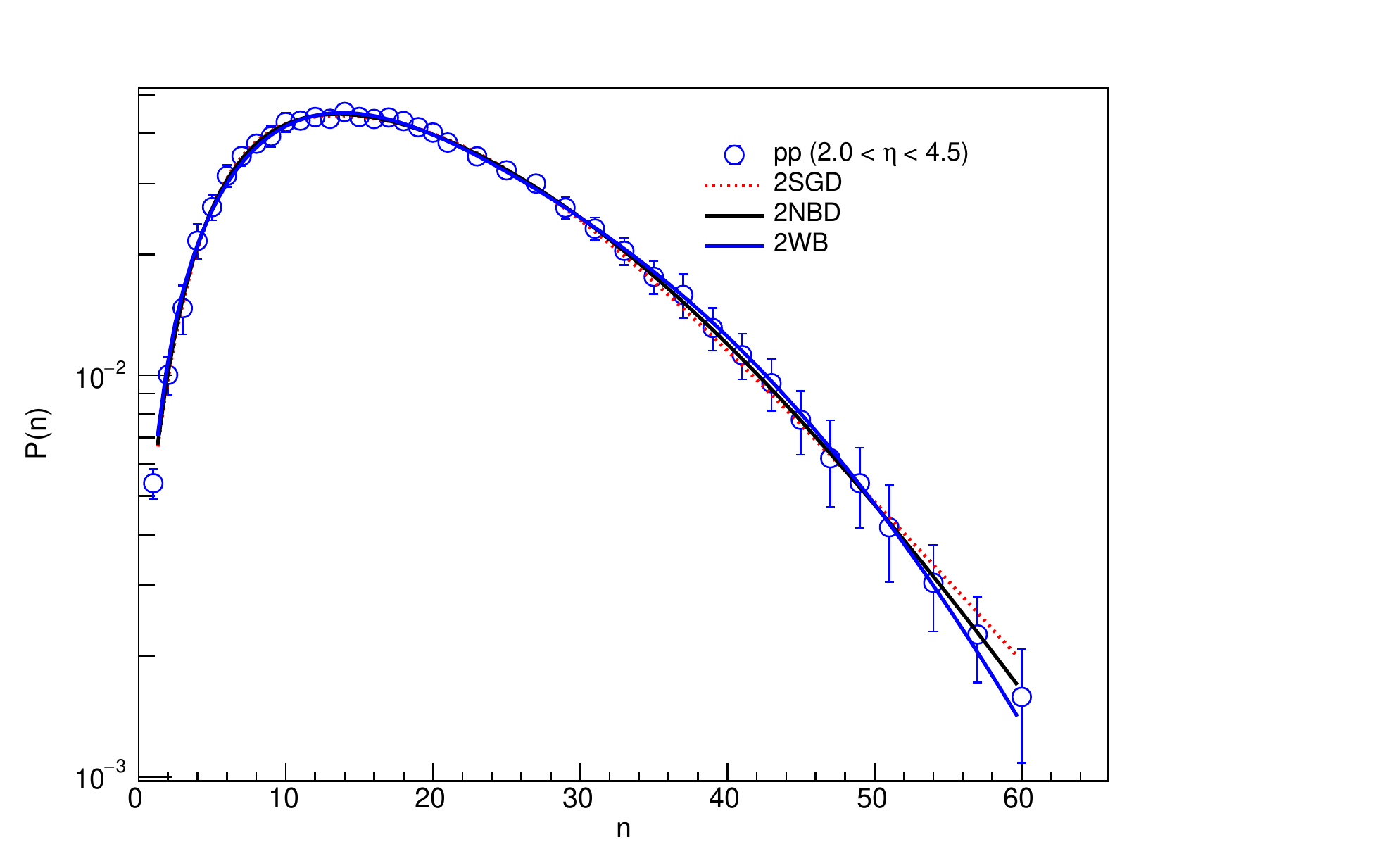}
\caption{Data on charged particle multiplicity distributions in $pp$ minimum bias and hard QCD events at $\sqrt{s}$= 7~TeV with fits from 2NBD, 2SGD, and 2WB(top to bottom) distributions in the forward pseudorapidity region ($\eta$= 2.0 to 4.5.)}
\end{figure}

\subsection{ MOMENTS OF MULTIPLICITY DISTRIBUTIONS}
The possibility of discovering correlations amongst the charged particles produced in collisions, higher-order moments and the cumulants are the precise tools \cite{Suz}.~The deviation w.r.t. independent and uncorrelated production of particles can be measured well using the factorial moments, $F_{q}$ \cite{IM}.\\
Figures~4-5 show for NBD, SGD and WB distributions, the normalized moments $C_{q}$ (equation(7)) for the two categories of events: minimum bias and hard-QCD events.~Similarly figures~6-7 show the normalized factorial moments $F_{q}$ (equation(8)).~The values for all the moments are given in Tables~IV and V.~Table~VI gives the values of the normalised moments $C_{q}$ and normalized factorial moments $F_{q}$ for the distributions 2NBD, 2SGD and 2WB.~Table IV summarize the values of normalized moments $C_{q}$ and normalized factorial moments $F_{q}$ with $q$ 2,3,4,5.~Following observations can be made;  (i) the values of moments, both $C_{q}$ and $F_{q}$ remain constant in different $\eta$ bins, with the exception of bin 4.0$<\eta<4.5$, in which the value is consistently lower for all moments.~Although the value is low, yet it agrees with the experimental value.   (ii) All values of moments $C_{q}$ and $F_{q}$  calculated from different distributions agree very well within the limits of error, for $q$~= 2,~3,~4 with the experimental values.~However the fit distributions overestimate the values of moments for $q$~=~5.~On comparison between the two categories of events, it is observed that the discrepencies between the fit values and data for $q$~=~5 moments are more pronounced for minimum bias events.\\  
The shape of the charged-particle multiplicity distribution analysed in terms of the $H_{q}$ shows quasi-oscillations.~As early as 1975, following the solution of QCD equations for the generating function, a special oscillation pattern for the ratio of cumulants to factorial moments $H_{q}=K_{q}/F_{q}$ was predicted with first minimum occuring around $q_{min}\backsim{5}$ and determined by inverse value of the QCD anomalous dimension;
\begin{equation}
\gamma_{0}=(2N_{c}\alpha_{s}/\pi)^{1/2}
\end{equation}
where $\alpha_{s}$ is the QCD running coupling constant, $N_{c}$~=~3 the number of colours.~Details can be found in references \cite{IM2,IM3,IM4,Cap,SLD}.~However this prediction was supposedly valid for the moments of parton multiplicity distributions, especially for gluons.~When the same analysis was done for final hadronic multiplicities in $e^{+}e^{-}$ and $pp/\overline{p}p$ experiments \cite{IM5}, similar oscillations in the $H_{q}$ ratio were observed.~In the present work, we study the behaviour of moments to check whether the multiplicity distributions in the forward region also follow the same trends.~We analyse both minimum bias events and hard QCD events using single as well as two component distributions.
  
Figures~8-9 show the ratio $H_{q}$ (equation(10)) as a function of the rank $q$ for the data, NBD, SGD and WB distributions, for the two categories of events: minimum bias and hard-QCD events for different pseudorapdity windows with $\Delta\eta$=0.5.~We find that the dependence of $H_{q}$ on $q$ is very similar in all the $\eta$ bins with a minimum value around $q$ = 6-7.~For minimum bias events, there is a disagreement between the data and the fit values at the highest $q$ values for all distributions.~But for the hard QCD events, the agreement between the data and the fit values is very good for WB and NBD, with SGD also following the data closely for all distributions.~There are slight discrepencies again towards the highest $q$ values in SGD.
  
In one of the studies by I.M. Dremin \cite{IM4}, it was pointed that in gluodynamics the gluon ratio $H_{q}$ has the minimum around $q\approx$4 or 5.~The quark factorial moments are larger than those of gluon jets.~First minimum of quark cumulants and of their ratio to factorial moments is positioned at $q\approx$8.~To translate theoretical predictions to experimentally measured values, implies transition from parton to particle level and hence puturbative QCD to non-purtaurbative QCD.~This involves implementation of some hadronisation models.~It is further argued that according to the hypothesis of local parton$-$hadron duality, the distributions of partons and hadrons differ by the numerical coefficient only which is determined by the partons recombined in a single hadron.~Therefore the normalized moments of gluons and quarks should be just related to the moments of the observed processes.~In a detailed study \cite{Gia} of the $e^{+}e^{-}$, $pp$ and $\overline{p}p$ in a wide range of energies, it has been concluded that the qualitative features of the behaviour of $H_{q}$ are very similar in all processes.~It is observed from the analysis of $\overline{p}p$ from UA5 collaboration, an abrupt descent and the subsequent oscillations are observed with minima at $q\approx$4 and 12 while the maxima  at $q\approx$9 and 15.

\begin{figure}
\includegraphics[width=3.7 in, height= 2.9 in]{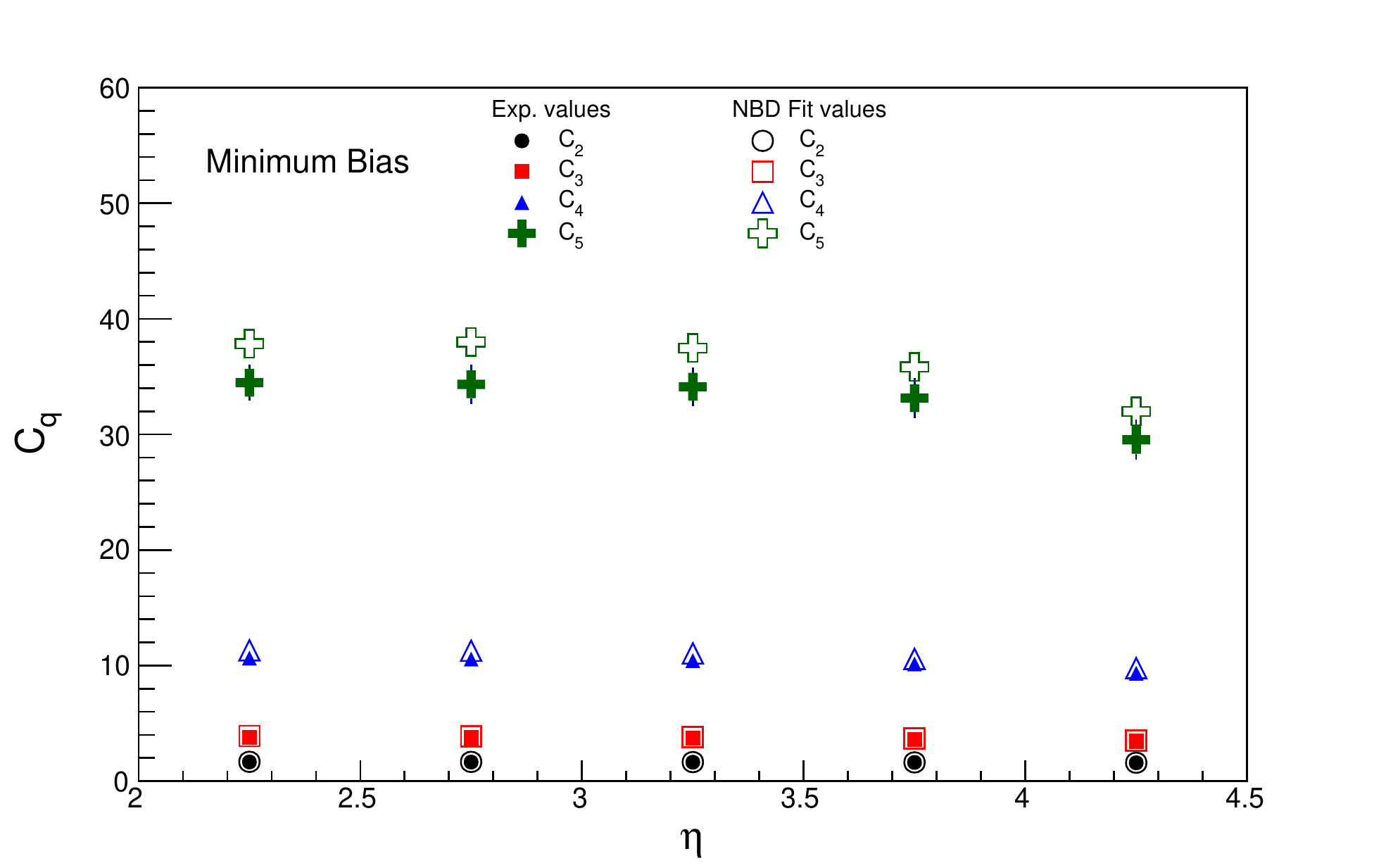}
\includegraphics[width=3.7 in, height= 2.9 in]{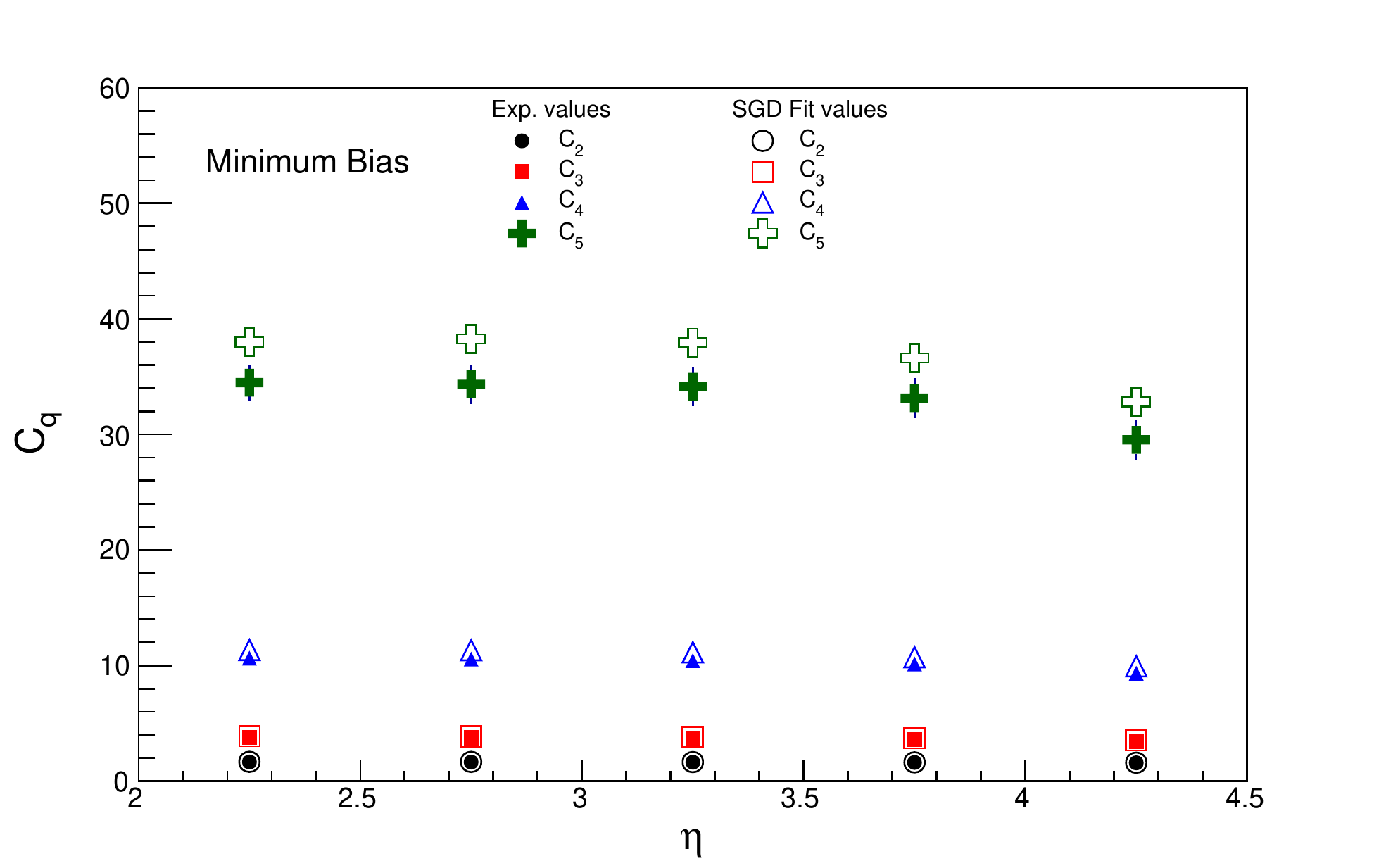}
\includegraphics[width=3.7 in, height= 2.9 in]{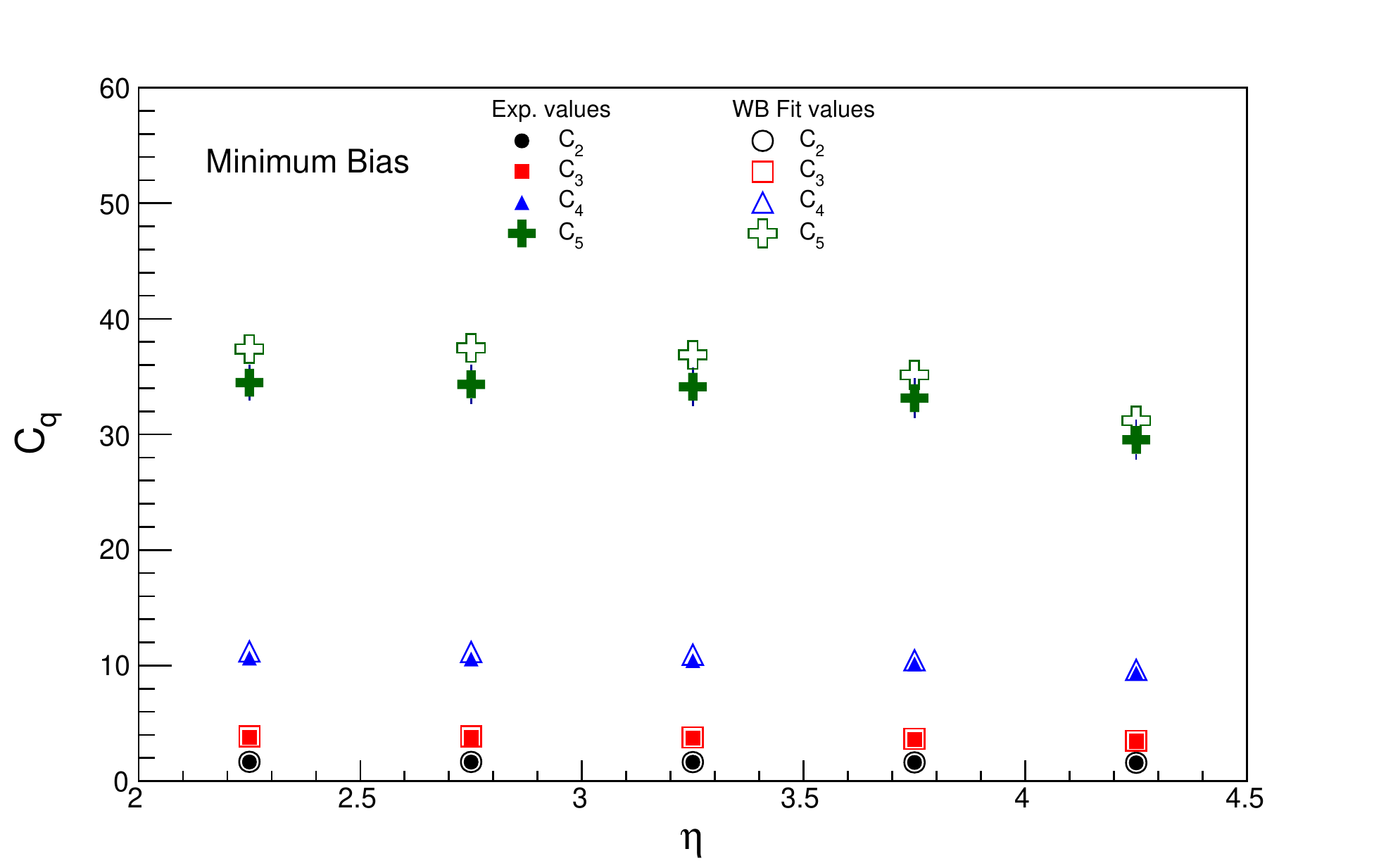}
\caption{Normalized moments of multiplicity distributions for minimum bias events in different pseudorapidity bins with bin size $\Delta\eta$=0.5.~Experimental values are shown in comparison to the fit values.}
\end{figure}

\begin{figure}
\includegraphics[width=3.7 in, height= 2.9 in]{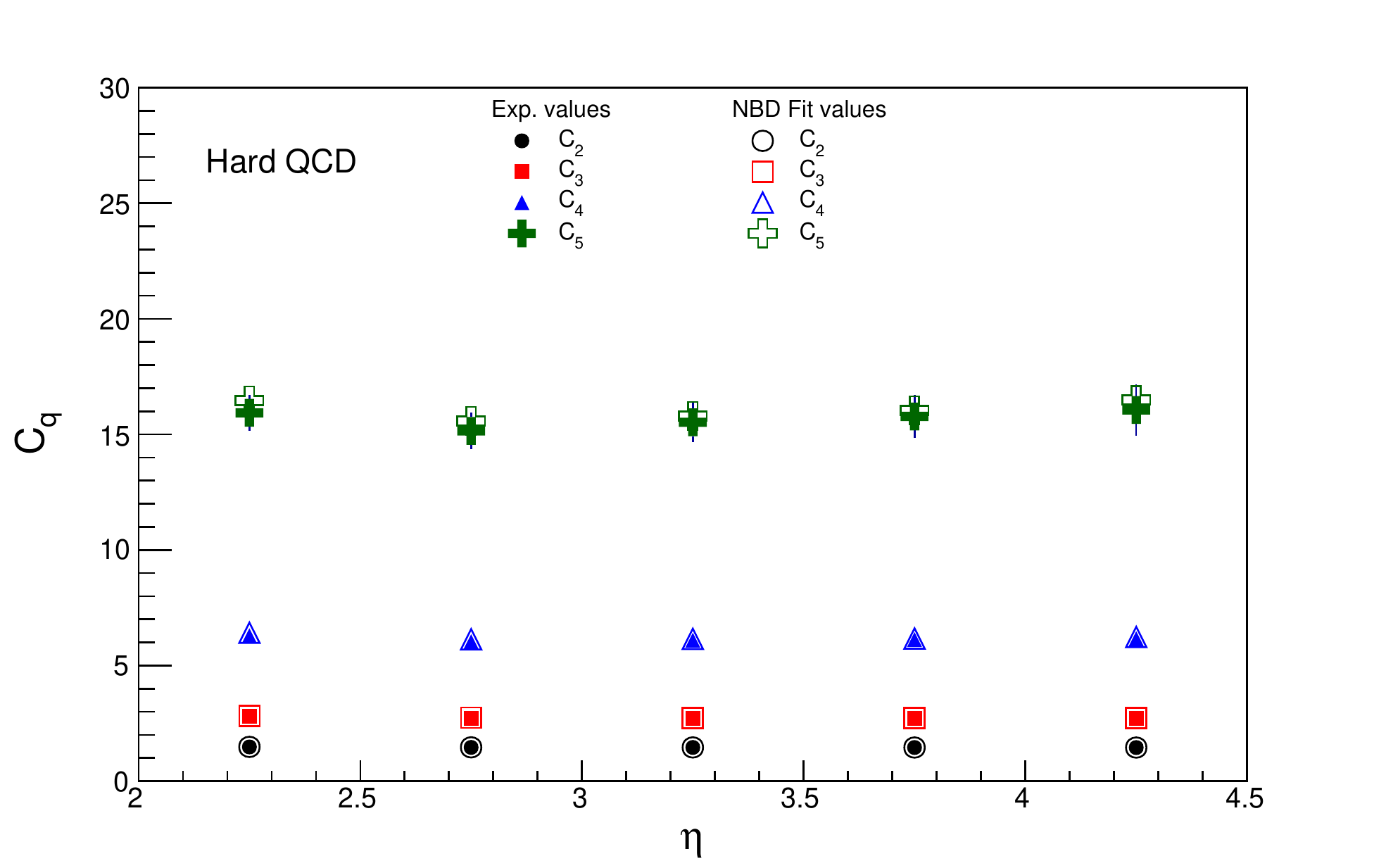}
\includegraphics[width=3.7 in, height= 2.9 in]{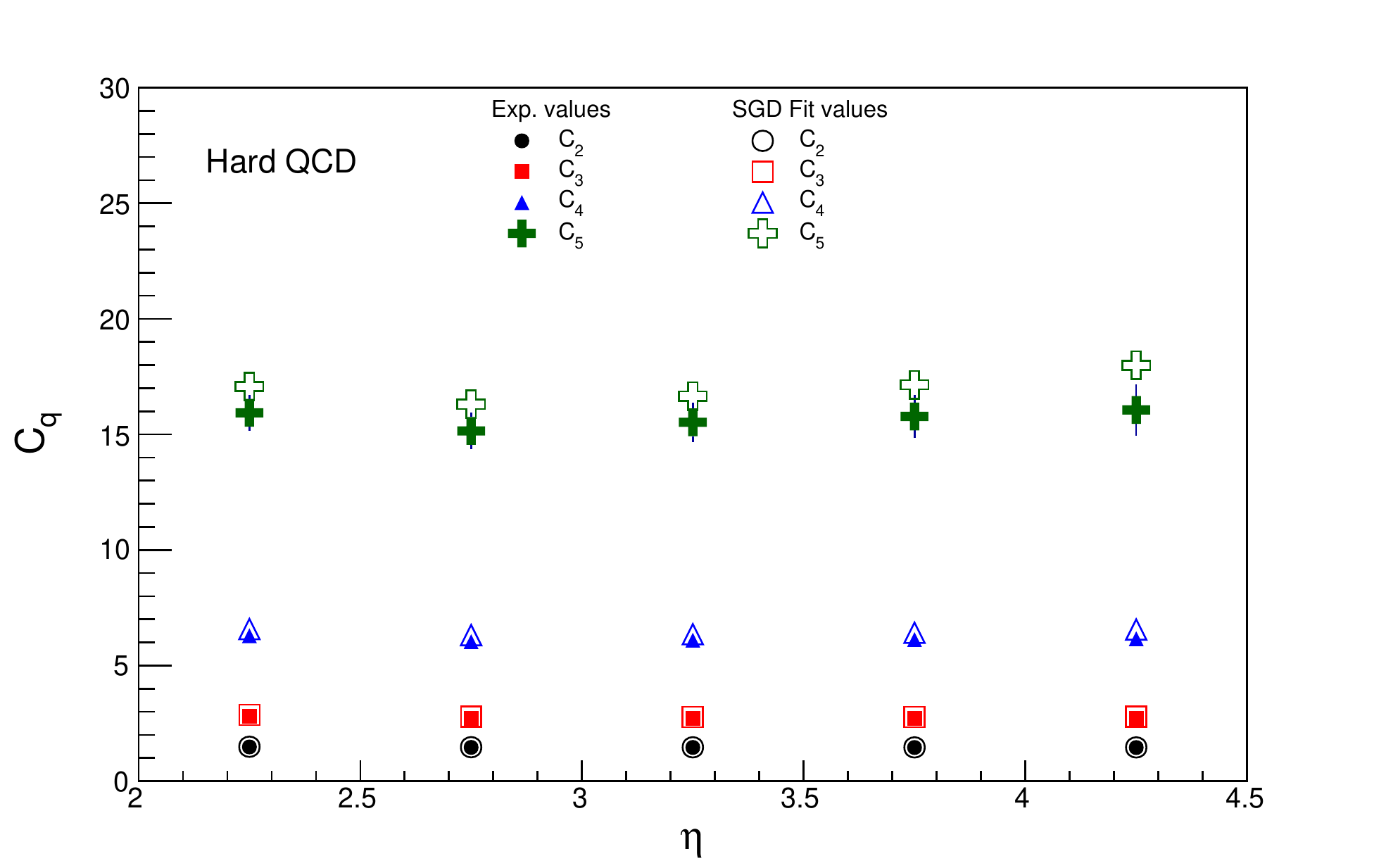}
\includegraphics[width=3.7 in, height= 2.9 in]{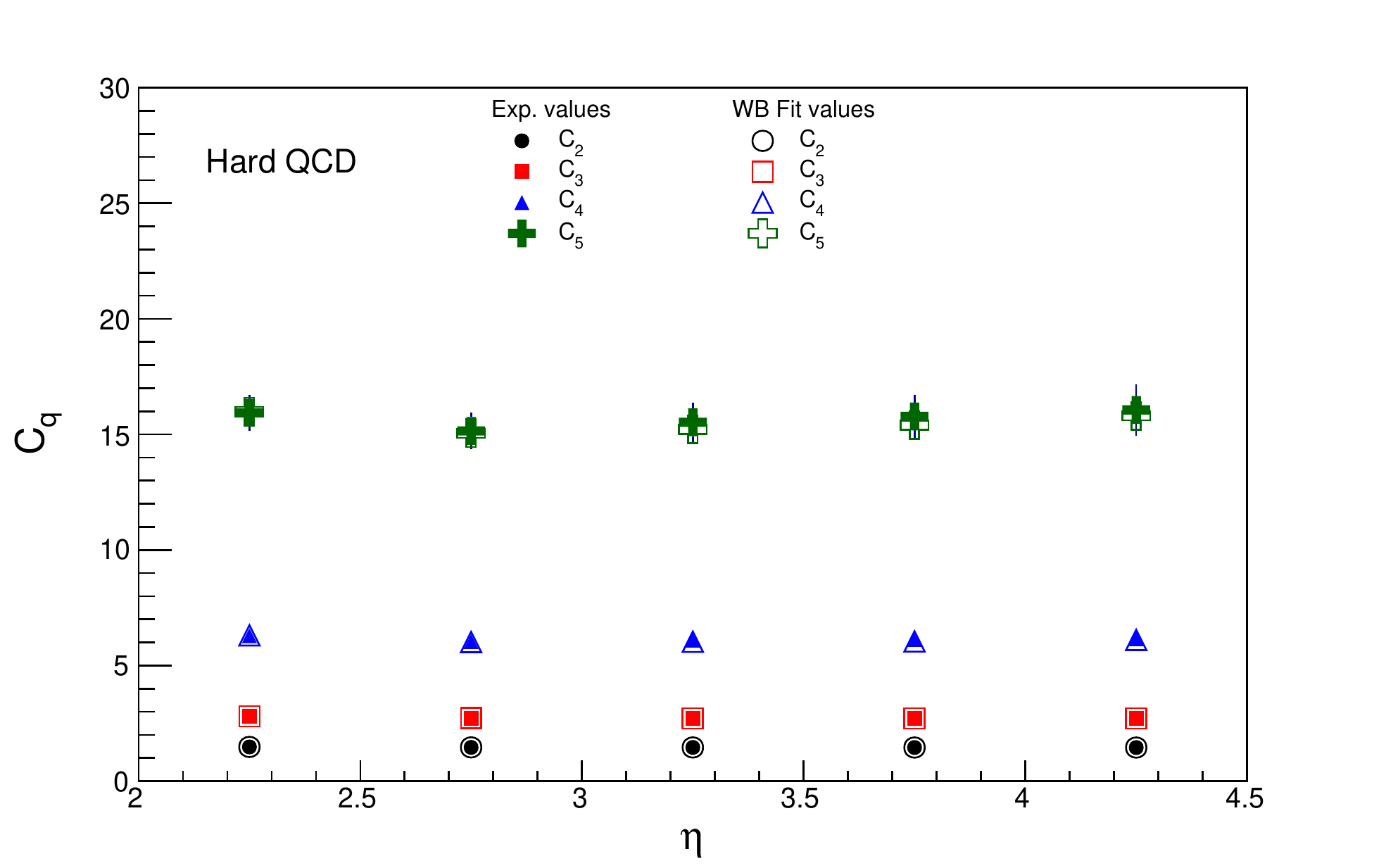}
\caption{Normalized moments of multiplicity distributions for hard-QCD events in different pseudorapidity bins with bin size $\Delta\eta$=0.5. Experimental values are shown in comparison to the fit values.}
\end{figure}

\begin{figure}
\includegraphics[width=3.7 in, height= 2.9 in]{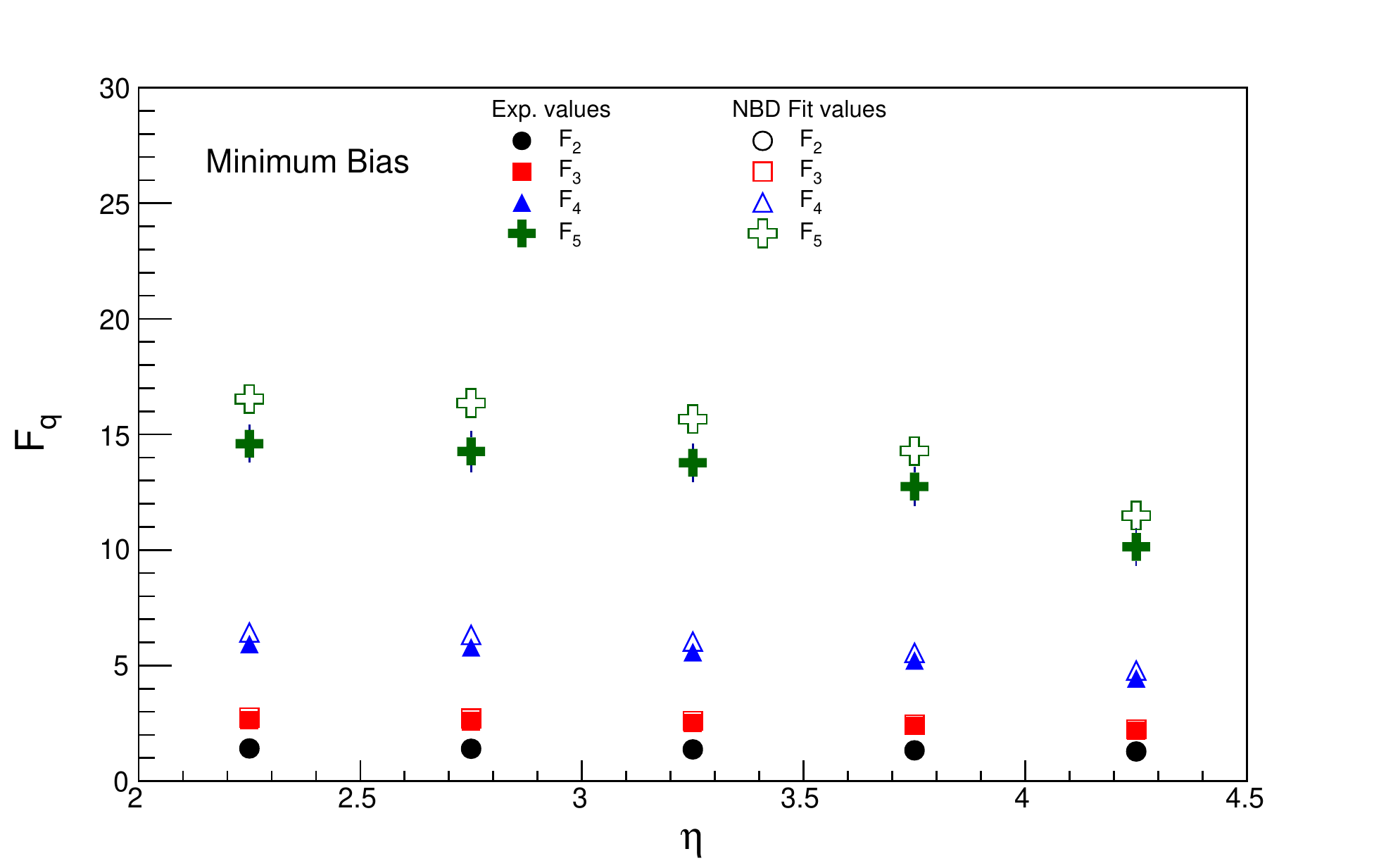}
\includegraphics[width=3.7 in, height= 2.9 in]{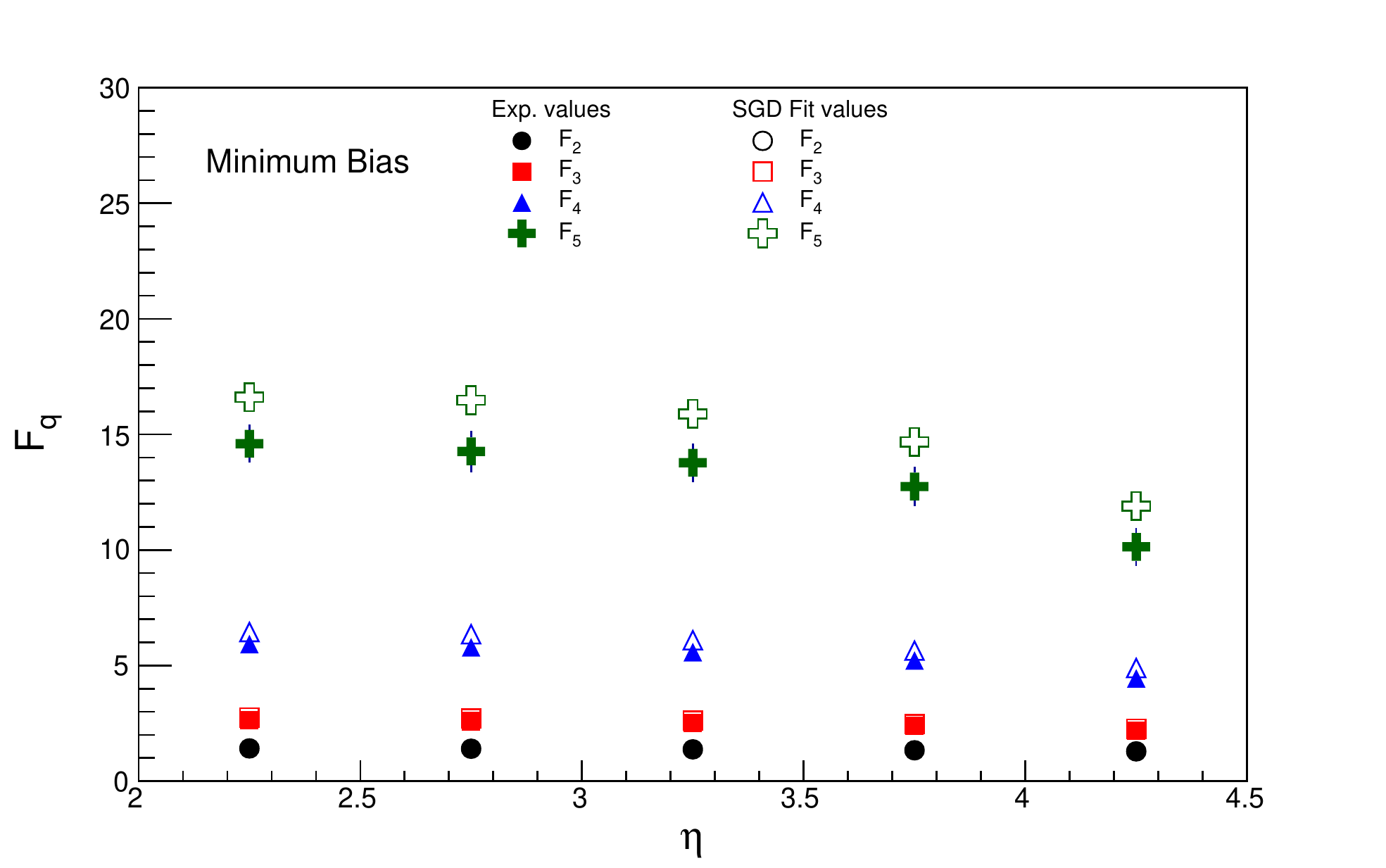}
\includegraphics[width=3.7 in, height= 2.9 in]{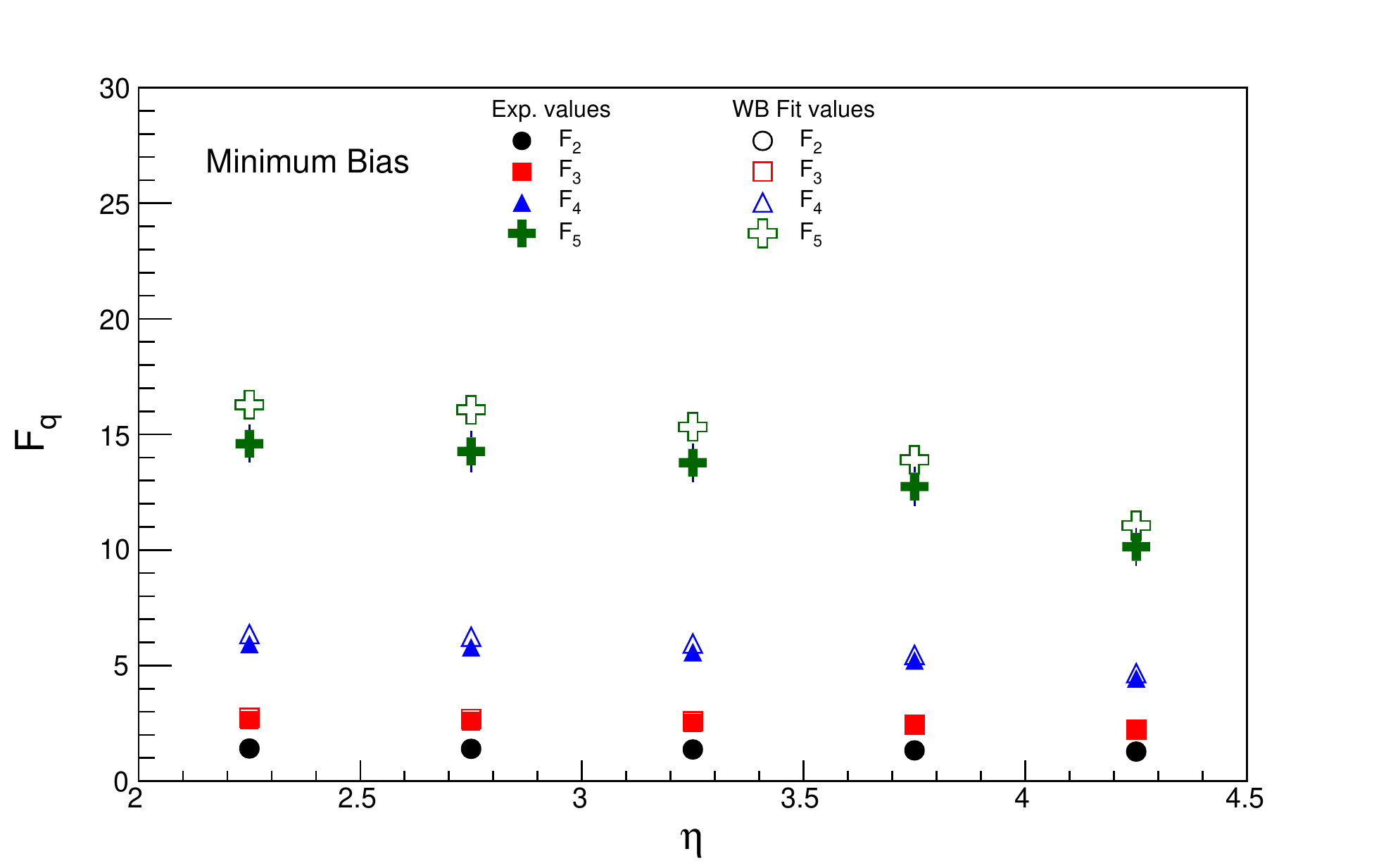}
\caption{Normalized factorial moments of multiplicity distributions for minimum bias events in different pseudorapidity bins with bin size $\Delta\eta$=0.5.~Experimental values are shown in comparison to the fit values.}
\end{figure}

\begin{figure}
\includegraphics[width=3.7 in, height= 2.9 in]{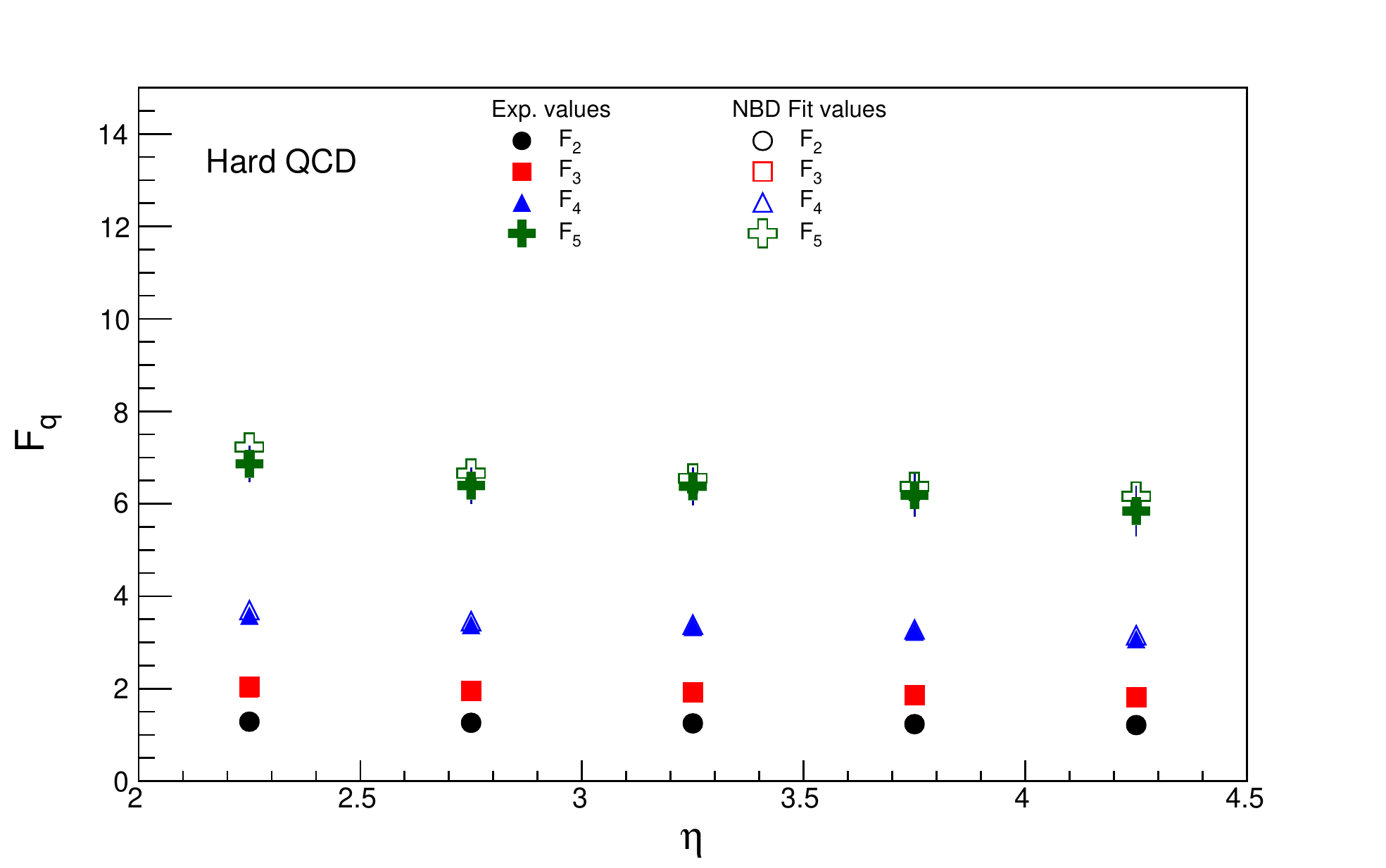}
\includegraphics[width=3.7 in, height= 2.9 in]{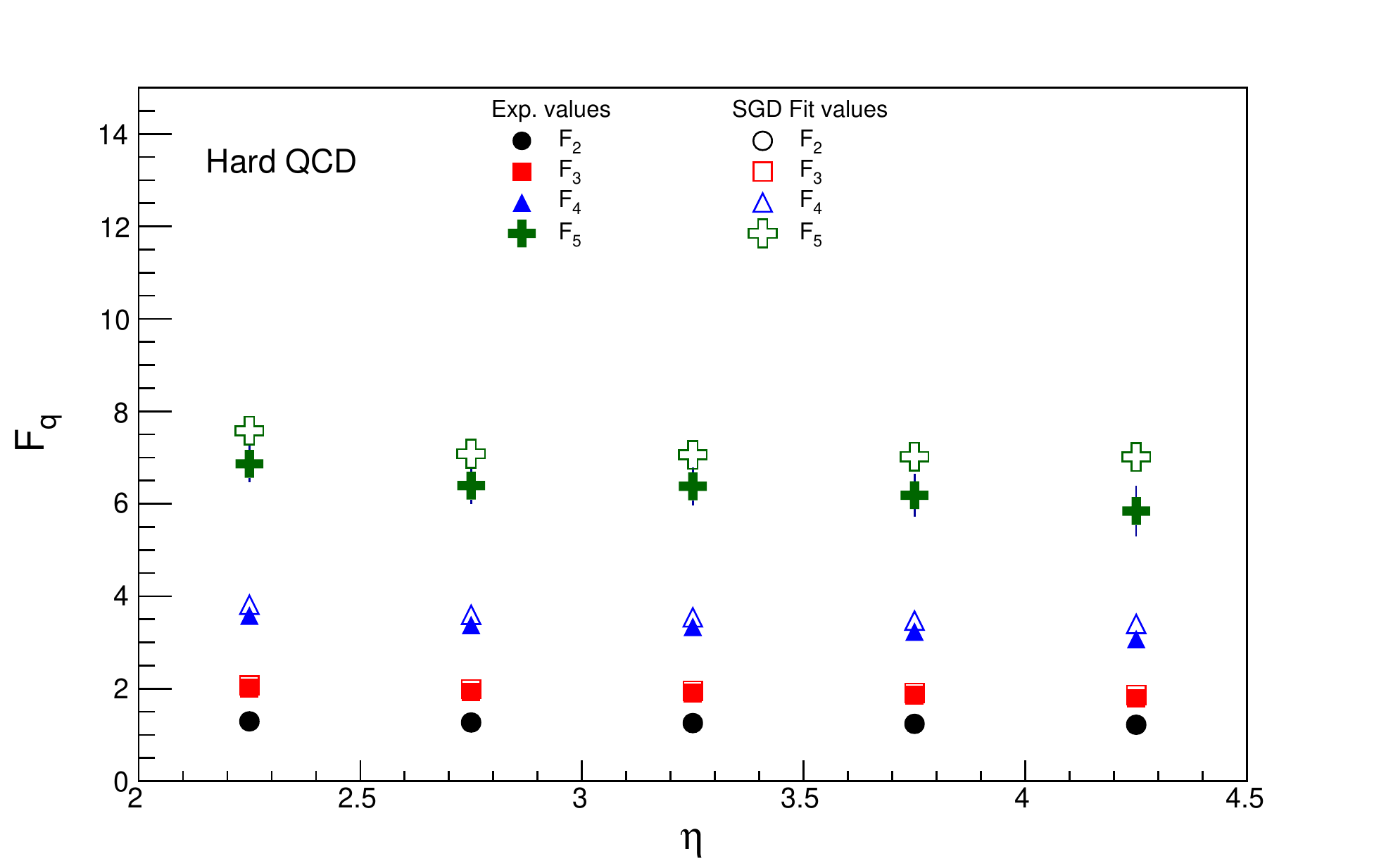}
\includegraphics[width=3.7 in, height= 2.9 in]{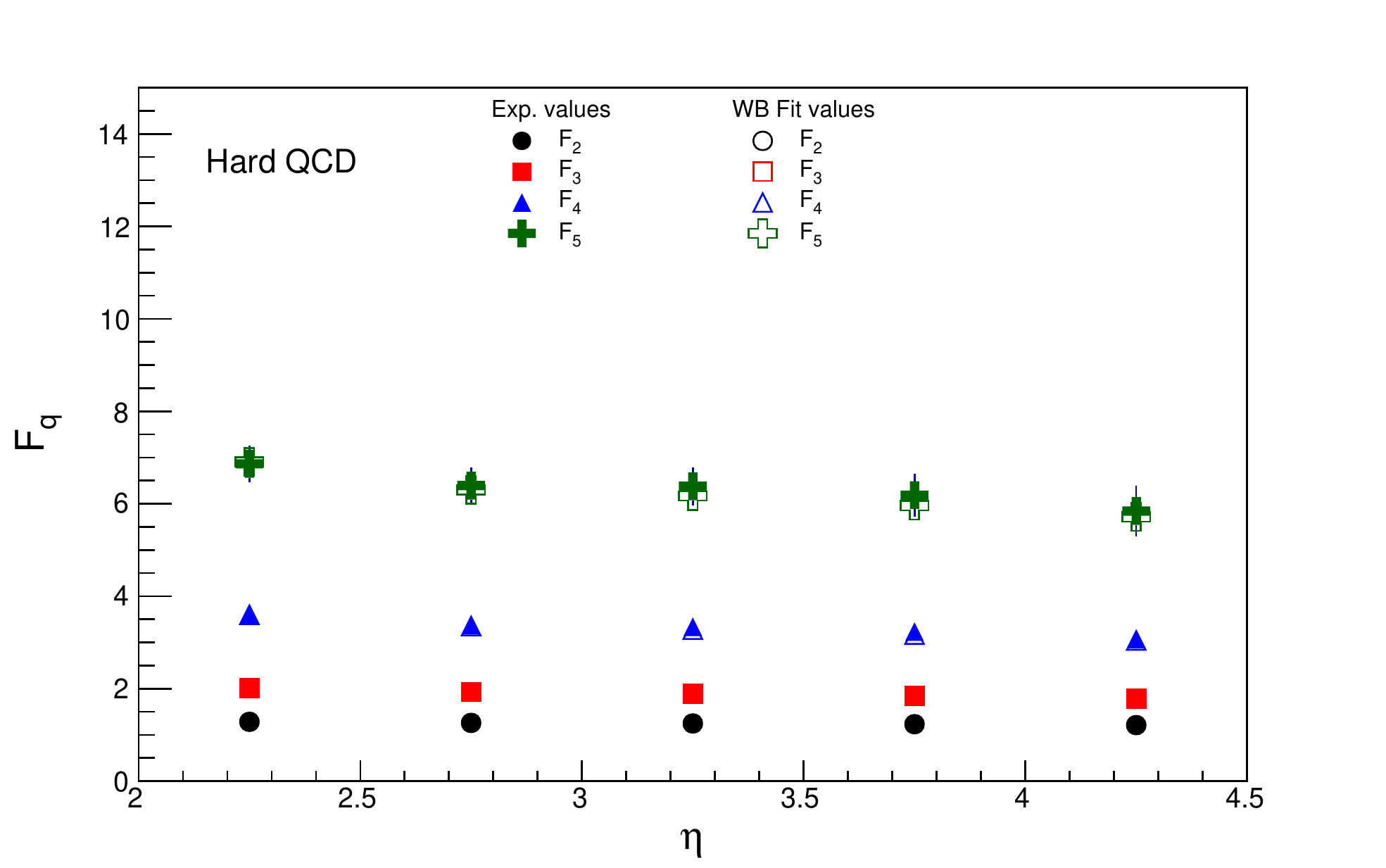}
\caption{Normalized factorial moments of multiplicity distributions for hard-QCD events in different pseudorapidity bins with bin size $\Delta\eta$=0.5.~Experimental values are shown in comparison to the fit values.}
\end{figure}

\begin{figure}
\includegraphics[width=3.9 in, height= 3.9 in]{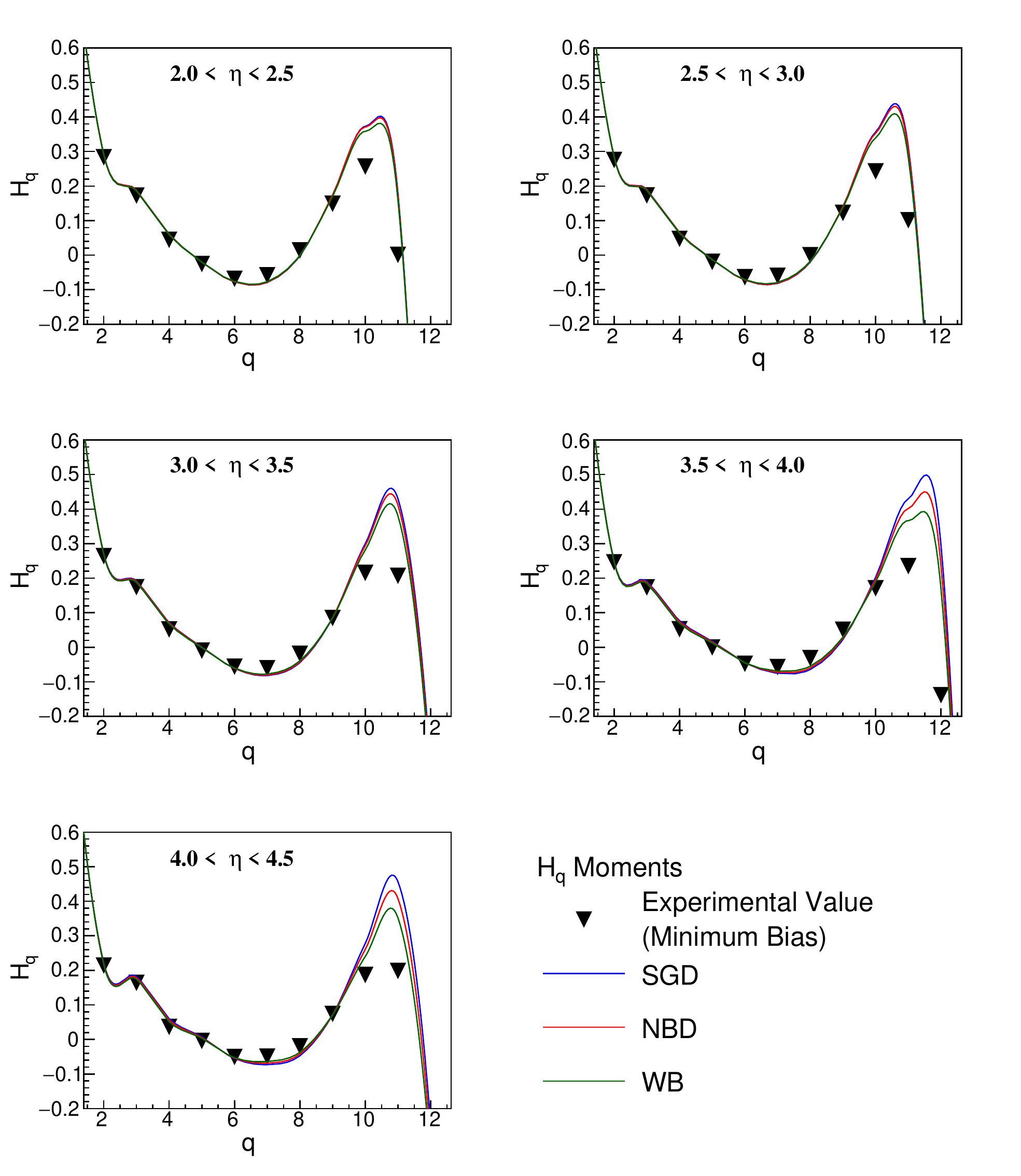}
\caption{Dependence of experimental values of $H_{q}$ moments on the rank $q$ in comparison to the values predicted by various distributions, in different $\eta$ windows for minimum bias events.}
\end{figure}

\begin{figure}
\includegraphics[width=3.9 in, height= 3.9 in]{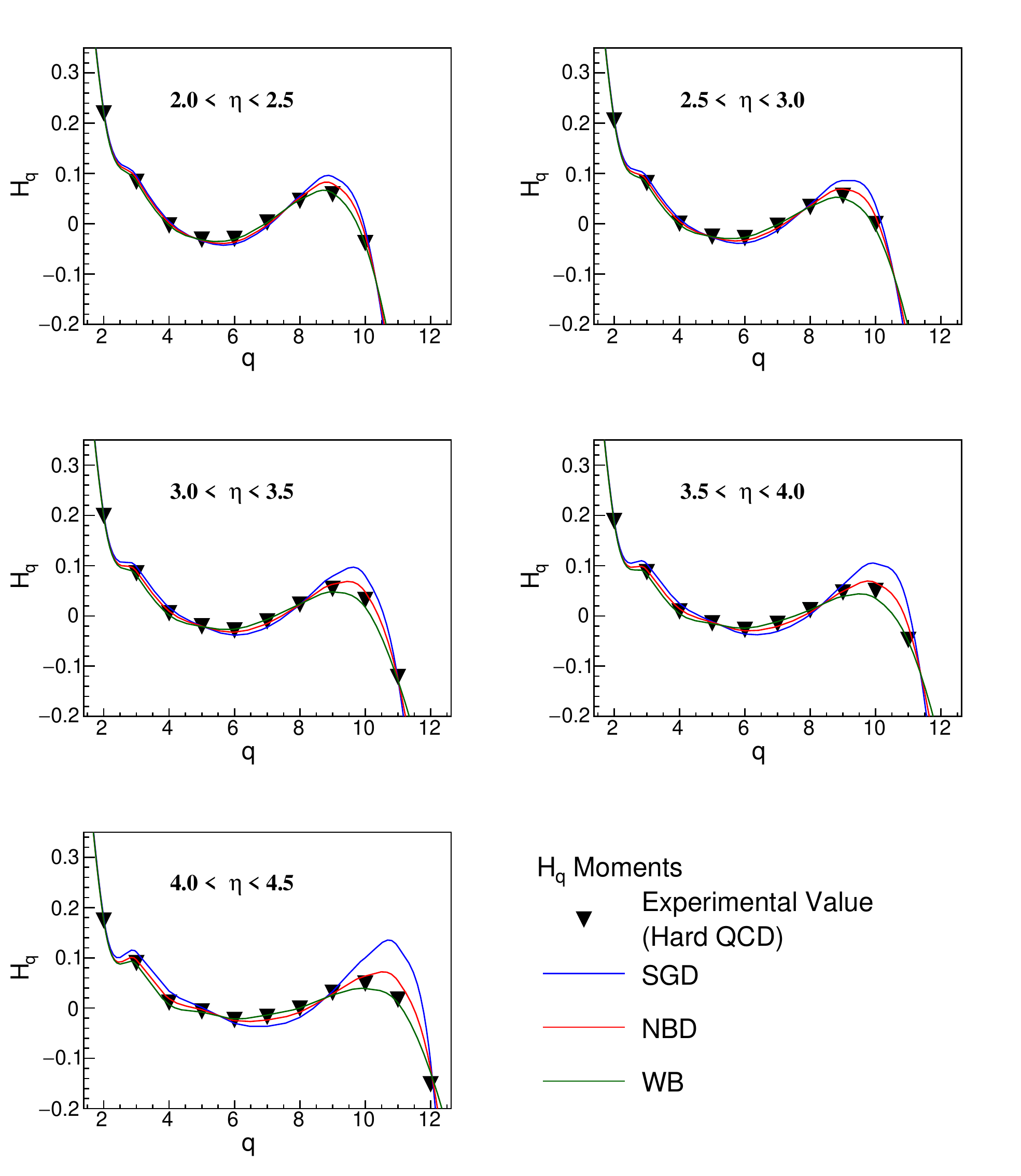}
\caption{Dependence of experimental values of $H_{q}$ moments on the rank $q$ in comparison to the values predicted by various distributions, in different $\eta$ windows for hard QCD events.}
\end{figure}
\begin{figure}
\includegraphics[width=3.3 in, height=3.7 in]{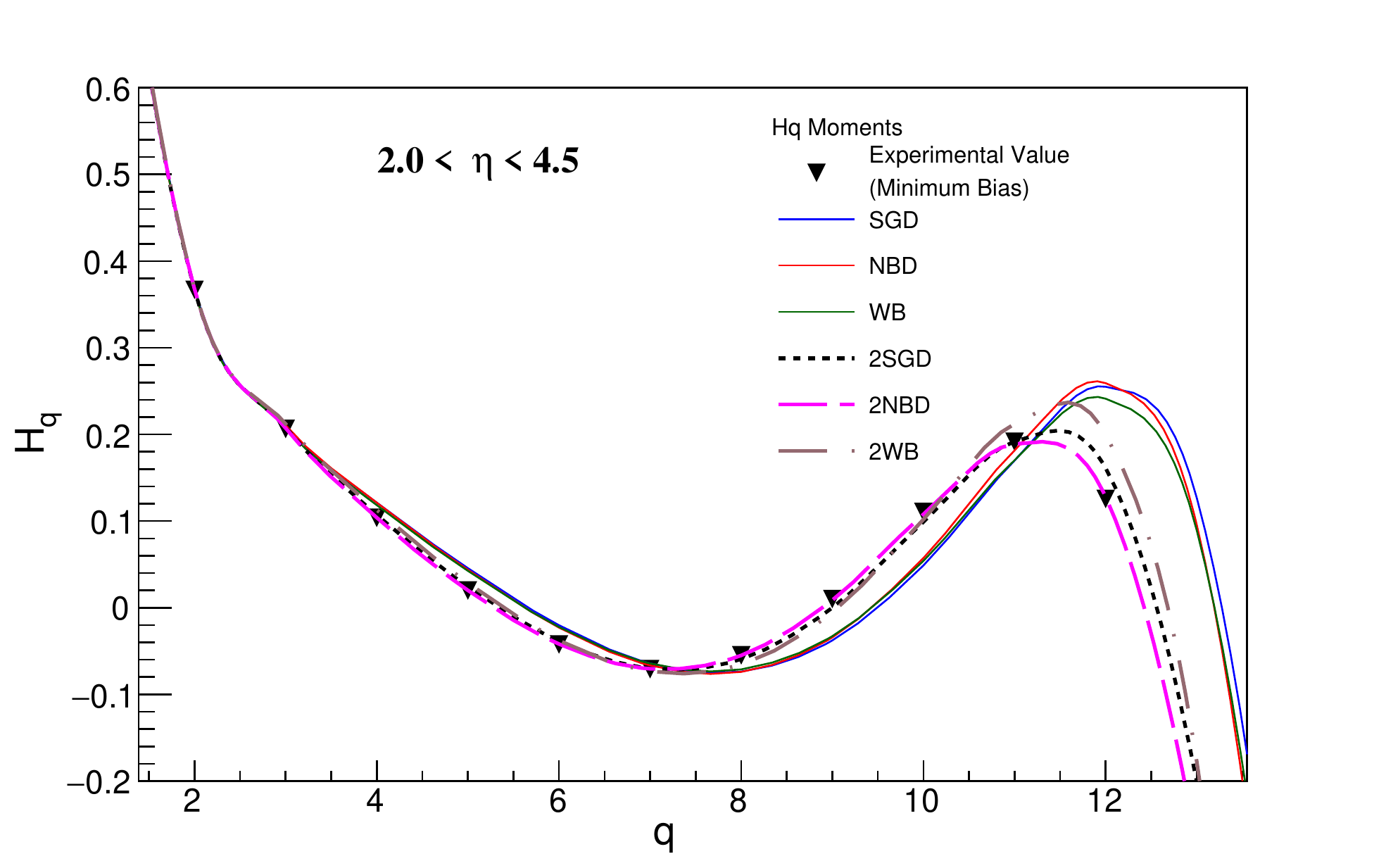}
\caption{Dependence of experimental values of $H_{q}$ moments on the rank $q$ in comparison to the values predicted by various distributions, in the forward region ($2.0<\eta<4.5$) for the minimum bias events.}
\end{figure}

\begin{figure}
\includegraphics[width=3.3 in, height= 3.7 in]{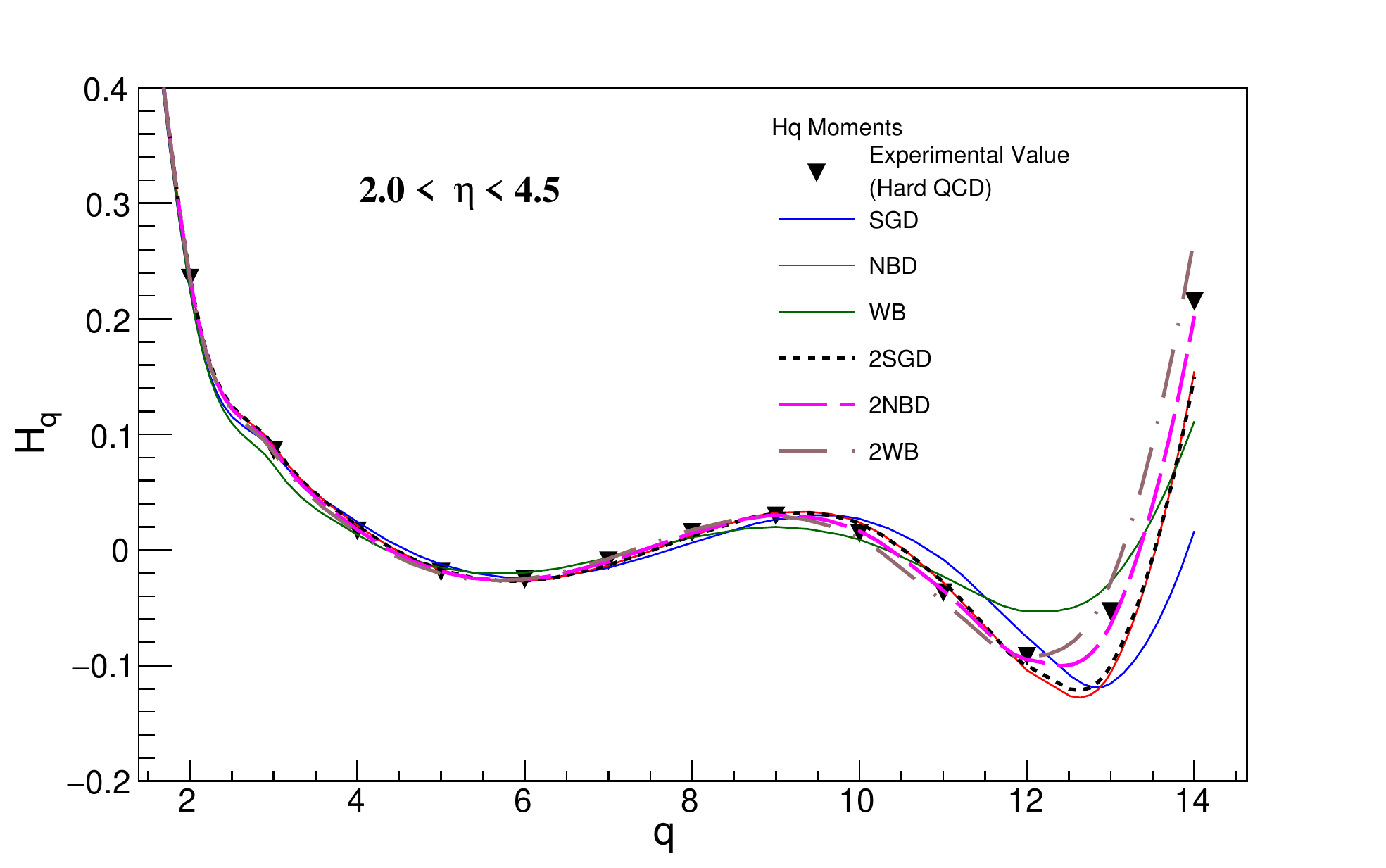}
\caption{Dependence of experimental values of $H_{q}$ moments on the rank $q$ in comparison to the values predicted by various distributions, in the forward region ($2.0<\eta<4.5$) for the hard QCD events.}
\end{figure}

Figures~10 shows the $H_{q}$ moments versus $q$ value calculated from the data and compared with NBD, SGD, WB, 2NBD, 2SGD, 2WB distributions, for the minimum bias events for the full forward region in the 2.0$<\eta<$4.5 interval.~2NBD followed by 2SGD best describe the data with minimum around $q\backsim$7. 
~Figure~11 shows the shape of the charged-particle multiplicity distribution of hard QCD events in full forward region, analyzed in terms of variation of $H_{q}$ moments as a function of $q$.~Comparison with NBD, SGD, WB, 2NBD, 2SGD, 2WB distributions shows that the 2NBD best describes the data, closely followed by 2WB and 2SGD.~The two minima appear at $q\approx$6 and 12 and the maxima appear at $q\backsim{9}$ and 15, with quasi-oscillations about zero for larger values of $q$.~These observations confirm the predictions from Quantum Chromodynamics and also the next-to-next-to-leading logarithm approximation (NNLLA) of perturbative QCD \cite{IM2,IM4,IM6}. 

Tables VII-VIII show the average charged particle multiplicity ($<n_{ch}>$) for the data in all pseudorapidity bins and full-forward region, in the two categories of events.~The average charged multiplicity is calculated from the probability distributions as, $<n_{ch}>=\sum nP(n)/\sum P(n)$.~Interesting observations reveal that the $<n_{ch}>$ in the $\eta$ intervals -2.5$< \eta <$-2.0 and 2.0$< \eta <$2.5 are very nearly the same within the error limits.~This indicates that the forward and backward regions are identical.~This result confirms the observation made in the paper by LHCb \cite{LHCb}.~Average multiplicity changes minimally over the $\eta$ intervals  for the minimum bias events. For the hard QCD events, it decreases from the bin 2.0$< \eta <$2.5 to 4.0$< \eta <$4.5 as expected due to the criterion of having atleast one track with $P_{T} >$1~GeV in each event.~Overall the values obtained from the fit values of the distributions NBD, SGD, WB, 2NBD, 2SGD, 2WB agree with the data values.~Finally, the $<n_{ch}>$ for the hard QCD events in the full forward region is larger than the minimum bias events.~But in each case the values agree with the fit values from all the distributions, with WB giving the closest agreement.

\normalsize
\section{CONCLUSION}
Comparison of multiplicity distributions at $\sqrt{s}$=7~TeV in restricted pseudorapidity ($\eta$) windows in the forward region obtained by the LHCb experiment is performed with three statistics inspired distributions namely negative binomial, shifted Gompertz and Weibull distributions.~Although the distributions fit the data very well in smaller $\eta$ windows (typically $\Delta\eta$~=~0.5), they all fail in the full forward region (2.0$<\eta<$4.5).~This kind of violation was observed with NBD at energies as low as 200$-$900~GeV \cite{Alba,Aln}.~A possible explanation of the effect was suggested by C.~Fuglesang \cite{Fug} in terms of purely phenomenological considerations indicating the presence of a substructure.~To overcome the violation, the multiplicity distribution was propsed to be a superposition of two component distributions.~Using this approach, we find that in the forward region, there is manifold reduction in the $\chi^{2}/ndf$ values and the distributions become statistically significant with a $p$-value corresponding to $CL>0.1\%$.

Shape of the charged$-$particle multiplicity distribution is related to the particle production.~To study the shape, normalized factorial moments are used.~If particles produced are correlated, the distribution is broader and the $F_{q}$ are greater than unity, if the particles are anticorrelated, the distribution becomes narrower reducing the $F_{q}$ values to less than unity.~The analysis of these moments, we find that values of $C_{q}$ and $F_{q}$ moments remain constant in different $\eta$ bins, with the exception of bin 4.0$<\eta<$4.5, in which the value is consistently lower for all moments.~This observation is consistent with the results from LHCb collaboration \cite{LHCb}.~Study of $H_{q}$  moments shows that dependence of $H_{q}$ on $q$ is very similar in all the $\eta$ bins with a minimum value around $q$ = 6-7.~For the hard QCD events, the agreement between the data and the fit values is very good for all distributions.~For minimum bias events, there is a disagreement between the data and the fit values at the highest $q$ values for all distributions.~The values of all $F_{q}$ are greater than unity, indicating the presence of correlations.

For the full forward region, 2.0$<\eta<$4.5, the $H_{q}$ moments versus $q$ analysis shows that the two component multiplicity distributions best describe the data for results obtained for hard QCD events.~In agreement with the QCD predictions, the two minima appear at $q\approx$6 and 12 and the maxima appear at $q\approx$9 and 15.~In particular, this observation also confirms the expectations of the next-to-next-to-leading logarithm approximation (NNLLA) from perturbative QCD.~The 2NBD best describes the data, closely followed by 2WB and 2SGD.~The distributions in forward and backward regions in the pseudorapidity 2.0$<|\eta|<$2.5 are found to be nearly identical.~The $<n_{ch}>$ for the hard QCD events in the full forward region is larger than the minimum bias events.~However in each case the values agree with the fit values from all the distributions, with 2WB giving the closest agreement.~The results obtained agree with the results from central region measured by CMS \cite{CMS}.
\begin{center}
{\bf ACKNOWLEDGMENT}
\end{center}	
The author R. Aggarwal is grateful to the DST, Government of India, for the INSPIRE faculty grant.



\begin{table*}[t]
\begin{tabular}{|c|c|c|c|c|c|} 
\hline 
 \multicolumn{5}{|c|}{NBD} \\ 
 \hline 
 $\eta$  & k  & $<n>$   & $\chi^2/ndf$ & p-value \\ 
 \hline 
 -2.5 $< \eta <$ -2.0  &     1.059 $\pm$    0.060 &     3.085 $\pm$    0.055&       2.79 / 15 &     1.00  \\ \hline 
2.0 $< \eta <$ 2.5  &     1.116 $\pm$    0.040 &     3.045 $\pm$    0.045&      14.16 / 17 &     0.66  \\ \hline 
2.5 $< \eta <$ 3.0  &     1.149 $\pm$    0.041 &     2.955 $\pm$    0.045&      14.52 / 17 &     0.63  \\ \hline 
3.0 $< \eta <$ 3.5  &     1.226 $\pm$    0.045 &     2.829 $\pm$    0.043&      12.63 / 17 &     0.76  \\ \hline 
3.5 $< \eta <$ 4.0  &     1.350 $\pm$    0.049 &     2.689 $\pm$    0.041&      15.83 / 17 &     0.54  \\ \hline 
4.0 $< \eta <$ 4.5  &     1.451 $\pm$    0.067 &     2.508 $\pm$    0.043&       8.18 / 14 &     0.88  \\ \hline 
2.0 $< \eta <$ 4.5  &     1.294 $\pm$    0.020 &    12.956 $\pm$    0.102&     213.82 / 36 &     $<$0.01  \\ \hline 
\hline 
 \multicolumn{5}{|c|}{SGD} \\ 
 \hline 
 $\eta$  & $\beta$   & $b$  & $\chi^2/ndf$ & p-value \\ 
 \hline 
-2.5 $< \eta <$ -2.0  &   0.061 $\pm$    0.058 &     0.292 $\pm$    0.009 &  2.70 / 15 &     1.00  \\ \hline 
2.0 $< \eta <$ 2.5  &     0.110 $\pm$    0.037&      0.305 $\pm$    0.005 & 	14.18 / 17 &     0.65  \\ \hline 
2.5 $< \eta <$ 3.0  &     0.137 $\pm$    0.038&      0.318 $\pm$    0.005 &	14.80 / 17 &     0.61  \\ \hline 
3.0 $< \eta <$ 3.5  &     0.200 $\pm$    0.039&      0.343 $\pm$    0.005 &     13.72 / 17 &     0.69  \\ \hline 
3.5 $< \eta <$ 4.0  &     0.300 $\pm$    0.042&      0.378 $\pm$    0.005 &     19.07 / 17 &     0.32  \\ \hline 
4.0 $< \eta <$ 4.5  &     0.370 $\pm$    0.052&      0.414 $\pm$    0.007 &	 10.48 / 14 &     0.73  \\ \hline 
2.0 $< \eta <$ 4.5  &     0.409 $\pm$    0.029&      0.092 $\pm$    0.002 &	329.70 / 36 &     $<$0.01  \\ \hline 
\hline 
 \multicolumn{5}{|c|}{WB} \\ 
 \hline 
 $\eta$  & $K$ & $\lambda$  & $\chi^2/ndf$ & p-value \\ 
 \hline 
-2.5 $< \eta <$ -2.0  &     1.021 $\pm$    0.019 &     3.569 $\pm$    0.057&       2.58 / 15 &     1.00  \\ \hline 
2.0 $< \eta <$ 2.5  &     1.040 $\pm$    0.012 &     3.536 $\pm$    0.047&      12.33 / 17 &     0.78  \\ \hline 
2.5 $< \eta <$ 3.0  &     1.049 $\pm$    0.012 &     3.446 $\pm$    0.047&      12.23 / 17 &     0.79  \\ \hline 
3.0 $< \eta <$ 3.5  &     1.068 $\pm$    0.012 &     3.319 $\pm$    0.044&       9.60 / 17 &     0.92  \\ \hline 
3.5 $< \eta <$ 4.0  &     1.097 $\pm$    0.012 &     3.176 $\pm$    0.042&      10.62 / 17 &     0.88  \\ \hline 
4.0 $< \eta <$ 4.5  &     1.120 $\pm$    0.015 &     2.991 $\pm$    0.044&       4.78 / 14 &     0.99  \\ \hline 
2.0 $< \eta <$ 4.5  &     1.125 $\pm$    0.008 &    13.662 $\pm$    0.103&     254.03 / 36 &     $<$0.01  \\ \hline 
\end{tabular} 
\caption{Fit parameters of three distributions for Minimum Bias Events} 
\end{table*} 

\begin{table*}[t]
\begin{tabular}{|c|c|c|c|c|c|} 

\hline 
 \multicolumn{5}{|c|}{NBD} \\ 
 \hline 
 $\eta$  & $k$  & $<n>$   & $\chi^2/ndf$ & p-value \\ 
 \hline 
 -2.5 $< \eta <$ -2.0  &     1.748 $\pm$    0.076 &     4.645 $\pm$    0.061&       6.34 / 16 &     0.98  \\ \hline 
2.0 $< \eta <$ 2.5  &     2.396 $\pm$    0.090 &     4.846 $\pm$    0.054&       6.91 / 17 &     0.98  \\ \hline 
2.5 $< \eta <$ 3.0  &     2.693 $\pm$    0.101 &     4.754 $\pm$    0.053&       2.42 / 17 &     1.00  \\ \hline 
3.0 $< \eta <$ 3.5  &     2.821 $\pm$    0.111 &     4.520 $\pm$    0.052&       1.58 / 17 &     1.00  \\ \hline 
3.5 $< \eta <$ 4.0  &     2.964 $\pm$    0.123 &     4.229 $\pm$    0.050&       2.47 / 17 &     1.00  \\ \hline 
4.0 $< \eta <$ 4.5  &     3.078 $\pm$    0.140 &     3.859 $\pm$    0.049&       6.63 / 17 &     0.99  \\ \hline 
2.0 $< \eta <$ 4.5  &     2.997 $\pm$    0.051 &    21.926 $\pm$    0.215&      12.77 / 36 &     1.00  \\ \hline
\hline   
 \multicolumn{5}{|c|}{SGD} \\ 
 \hline 
 $\eta$  & $\beta$  & $b$  & $\chi^2/ndf$ & p-value \\ 
 \hline 
-2.5 $< \eta <$ -2.0  &   0.732 $\pm$    0.063&      0.273 $\pm$    0.006 & 	   4.69 / 16 &     1.00  \\ \hline 
2.0 $< \eta <$ 2.5  &     1.266 $\pm$    0.071&      0.304 $\pm$    0.004 & 	  13.46 / 17 &     0.71  \\ \hline 
2.5 $< \eta <$ 3.0  &     1.493 $\pm$    0.076&      0.326 $\pm$    0.004 &	  10.31 / 17 &     0.89  \\ \hline 
3.0 $< \eta <$ 3.5  &     1.570 $\pm$    0.081&      0.347 $\pm$    0.005 &      9.56 / 17 &     0.92  \\ \hline 
3.5 $< \eta <$ 4.0  &     1.636 $\pm$    0.086&      0.374 $\pm$    0.005 &     15.78 / 17 &     0.54  \\ \hline 
4.0 $< \eta <$ 4.5  &     1.649 $\pm$    0.092&      0.409 $\pm$    0.005 &     34.36 / 17 &     0.01  \\ \hline 
2.0 $< \eta <$ 4.5  &     3.202 $\pm$    0.059&      0.095 $\pm$    0.001 &     79.50 / 36 &     $<$0.01  \\ \hline  
 \hline 
 \multicolumn{5}{|c|}{WB} \\ 
 \hline 
 $\eta$  & $K$   & $\lambda$  & $\chi^2/ndf$ & p-value \\ 
 \hline 
-2.5 $< \eta <$ -2.0  &     1.239 $\pm$    0.019 &     5.196 $\pm$    0.063&       3.23 / 16 &     1.00  \\ \hline 
2.0 $< \eta <$ 2.5  &     1.368 $\pm$    0.017 &     5.452 $\pm$    0.058&       1.32 / 17 &     1.00  \\ \hline 
2.5 $< \eta <$ 3.0  &     1.413 $\pm$    0.017 &     5.367 $\pm$    0.057&       2.63 / 17 &     1.00  \\ \hline 
3.0 $< \eta <$ 3.5  &     1.421 $\pm$    0.018 &     5.123 $\pm$    0.056&       3.86 / 17 &     1.00  \\ \hline 
3.5 $< \eta <$ 4.0  &     1.426 $\pm$    0.018 &     4.816 $\pm$    0.054&       2.42 / 17 &     1.00  \\ \hline 
4.0 $< \eta <$ 4.5  &     1.420 $\pm$    0.019 &     4.414 $\pm$    0.052&       1.95 / 17 &     1.00  \\ \hline 
2.0 $< \eta <$ 4.5  &     1.826 $\pm$    0.014 &    23.565 $\pm$    0.231&      64.72 / 36 &     $<$0.01  \\ \hline 
\end{tabular} 
\caption{Fit parameters of three distributions for hard QCD Events} 
\end{table*} 

\begin{table*}[t]
\begin{tabular}{|c||c|c|c|c|c|c|c|} 
\hline 
 \\[-1em] 
 & \multicolumn{7}{c|}{2 NBD}\\ 
 \hline
 Events   &  $k1$ & $ <n1>$  & $\alpha$  &  $k2$ & $ <n2>$ & $\chi^2/ndf$ & p-value \\ 
 \hline   
Minimum Bias  &     2.157 $\pm$    0.161 &     7.595 $\pm$    0.845  &     0.623 $\pm$    0.086  &     4.398 $\pm$    1.139  &    24.068 $\pm$    2.238 &      14.90 / 33 &     1.00  \\ \hline 
Hard QCD  &     3.754 $\pm$    0.608 &    13.321 $\pm$    7.313  &     0.402 $\pm$    0.663  &     5.502 $\pm$    3.940  &    27.277 $\pm$    8.259 &       4.41 / 33 &     1.00  \\ \hline 
 \\[-1em] 
 & \multicolumn{7}{c|}{2 SGD}\\ 
 \hline
 Events   & $\beta1$   &  $b1$ &  $\alpha$  & $\beta2$ & $b2 $ & $\chi^2/ndf$ & p-value \\ 
 \hline   
Minimum Bias  & 1.362 $\pm$    0.072  &     0.198 $\pm$    0.015 &     0.640 $\pm$    0.057  &     5.434 $\pm$    1.552 &      0.098 $\pm$    0.004  &      7.75 / 33 &     1.00  \\ \hline 
Hard QCD  &     3.572 $\pm$    0.194  &     0.184 $\pm$    0.040 &     0.285 $\pm$    0.204  &     5.490 $\pm$    2.720 &      0.094 $\pm$    0.004  &      6.91 / 33 &     1.00  \\ \hline 
 \\[-1em] 
 &\multicolumn{7}{c|}{2 WB}\\ 
 \hline
 Events   &   $K1$  & $\lambda1$ & $\alpha$ &  $K2 $&  $\lambda2$& $\chi^2/ndf$ & p-value \\ 
 \hline   
Minimum Bias  &     2.248 $\pm$    0.236 &     6.797 $\pm$    0.222  &     0.086 $\pm$    0.018  &     1.171 $\pm$    0.018  &    15.728 $\pm$    0.314 &       6.84 / 33 &     1.00  \\ \hline 
Hard QCD  &     2.023 $\pm$    0.143 &    14.736 $\pm$    1.362  &     0.258 $\pm$    0.150  &     2.001 $\pm$    0.181  &    27.644 $\pm$    2.311 &       6.42 / 33 &     1.00  \\ \hline 
\end{tabular} 
\caption{Fit parameter values, $\chi^2/ndf$ and $p$-values obtained for the minimum bias and hard QCD events from 2NBD, 2SGD and 2WB models. } 
\end{table*} 

\begin{table*}[t]
\begin{tabular}{|c|c|c|c|c|c|c|c|c|} 
\hline 
$\eta$ &  $C_2$ & $C_3$  & $C_4$ & $C_5$ &  $F_2$ & $F_3$  & $F_4$ & $F_5$ \\ 
 \hline 
\hline 
 \multicolumn{9}{|c|}{Experimental values (Minimum Bias)}\\ 
 \hline-2.5 $< \eta <$ -2.0 &      1.648 $\pm$    0.029&       3.72 $\pm$     0.14&      10.19 $\pm$     0.63&      31.60 $\pm$     2.75&      1.393 $\pm$    0.027&       2.59 $\pm$     0.11 &      5.57 $\pm$     0.41 &     12.79 $\pm$     1.38 \\ \hline 
2.0 $< \eta <$ 2.5 &      1.652 $\pm$    0.013&       3.78 $\pm$     0.07&      10.65 $\pm$     0.33&      34.48 $\pm$     1.56&      1.395 $\pm$    0.012&       2.64 $\pm$     0.05 &      5.93 $\pm$     0.22 &     14.60 $\pm$     0.82 \\ \hline 
2.5 $< \eta <$ 3.0 &      1.644 $\pm$    0.013&       3.75 $\pm$     0.07&      10.56 $\pm$     0.35&      34.33 $\pm$     1.69&      1.381 $\pm$    0.012&       2.59 $\pm$     0.06 &      5.79 $\pm$     0.24 &     14.26 $\pm$     0.89 \\ \hline 
3.0 $< \eta <$ 3.5 &      1.632 $\pm$    0.014&       3.70 $\pm$     0.07&      10.43 $\pm$     0.34&      34.12 $\pm$     1.66&      1.359 $\pm$    0.013&       2.51 $\pm$     0.06 &      5.58 $\pm$     0.22 &     13.78 $\pm$     0.84 \\ \hline 
3.5 $< \eta <$ 4.0 &      1.614 $\pm$    0.014&       3.62 $\pm$     0.07&      10.13 $\pm$     0.35&      33.14 $\pm$     1.73&      1.327 $\pm$    0.013&       2.40 $\pm$     0.06 &      5.22 $\pm$     0.23 &     12.75 $\pm$     0.85 \\ \hline 
4.0 $< \eta <$ 4.5 &      1.580 $\pm$    0.015&       3.44 $\pm$     0.07&       9.33 $\pm$     0.35&      29.53 $\pm$     1.74&      1.272 $\pm$    0.013&       2.17 $\pm$     0.06 &      4.44 $\pm$     0.22 &     10.14 $\pm$     0.83 \\ \hline 
2.0 $< \eta <$ 4.5 &      1.665 $\pm$    0.009&       3.86 $\pm$     0.05&      11.15 $\pm$     0.25&      37.06 $\pm$     1.20&      1.579 $\pm$    0.009&       3.45 $\pm$     0.05 &      9.29 $\pm$     0.22 &     28.45 $\pm$     0.99 \\ \hline 
\hline 
 \multicolumn{9}{|c|}{NBD Fit values}\\ 
 \hline-2.5 $< \eta <$ -2.0 &      1.656 $\pm$    0.009&       3.78 $\pm$     0.04&      10.53 $\pm$     0.17&      33.24 $\pm$     0.71&      1.402 $\pm$    0.008&       2.65 $\pm$     0.03 &      5.84 $\pm$     0.10 &     13.79 $\pm$     0.32 \\ \hline 
2.0 $< \eta <$ 2.5 &      1.667 $\pm$    0.006&       3.90 $\pm$     0.03&      11.31 $\pm$     0.12&      37.84 $\pm$     0.53&      1.412 $\pm$    0.005&       2.75 $\pm$     0.02 &      6.43 $\pm$     0.07 &     16.53 $\pm$     0.26 \\ \hline 
2.5 $< \eta <$ 3.0 &      1.660 $\pm$    0.006&       3.88 $\pm$     0.03&      11.28 $\pm$     0.12&      38.00 $\pm$     0.56&      1.398 $\pm$    0.006&       2.71 $\pm$     0.02 &      6.32 $\pm$     0.07 &     16.36 $\pm$     0.27 \\ \hline 
3.0 $< \eta <$ 3.5 &      1.643 $\pm$    0.007&       3.80 $\pm$     0.03&      11.05 $\pm$     0.13&      37.47 $\pm$     0.58&      1.370 $\pm$    0.006&       2.60 $\pm$     0.02 &      6.04 $\pm$     0.08 &     15.66 $\pm$     0.28 \\ \hline 
3.5 $< \eta <$ 4.0 &      1.617 $\pm$    0.007&       3.68 $\pm$     0.03&      10.57 $\pm$     0.13&      35.83 $\pm$     0.59&      1.329 $\pm$    0.006&       2.45 $\pm$     0.02 &      5.55 $\pm$     0.07 &     14.28 $\pm$     0.27 \\ \hline 
4.0 $< \eta <$ 4.5 &      1.587 $\pm$    0.008&       3.51 $\pm$     0.03&       9.78 $\pm$     0.14&      31.99 $\pm$     0.63&      1.280 $\pm$    0.007&       2.24 $\pm$     0.02 &      4.78 $\pm$     0.08 &     11.49 $\pm$     0.27 \\ \hline 
2.0 $< \eta <$ 4.5 &      1.668 $\pm$    0.002&       3.90 $\pm$     0.01&      11.51 $\pm$     0.06&      39.74 $\pm$     0.31&      1.579 $\pm$    0.002&       3.47 $\pm$     0.01 &      9.56 $\pm$     0.05 &     30.46 $\pm$     0.25 \\ \hline 
\hline 
 \multicolumn{9}{|c|}{SGD Fit values}\\ 
 \hline-2.5 $< \eta <$ -2.0 &      1.656 $\pm$    0.009&       3.78 $\pm$     0.04&      10.52 $\pm$     0.17&      33.22 $\pm$     0.70&      1.402 $\pm$    0.008&       2.65 $\pm$     0.03 &      5.84 $\pm$     0.10 &     13.77 $\pm$     0.32 \\ \hline 
2.0 $< \eta <$ 2.5 &      1.669 $\pm$    0.006&       3.91 $\pm$     0.03&      11.35 $\pm$     0.11&      38.01 $\pm$     0.53&      1.413 $\pm$    0.005&       2.76 $\pm$     0.02 &      6.45 $\pm$     0.07 &     16.61 $\pm$     0.26 \\ \hline 
2.5 $< \eta <$ 3.0 &      1.662 $\pm$    0.006&       3.89 $\pm$     0.03&      11.33 $\pm$     0.12&      38.26 $\pm$     0.55&      1.399 $\pm$    0.006&       2.71 $\pm$     0.02 &      6.35 $\pm$     0.07 &     16.48 $\pm$     0.27 \\ \hline 
3.0 $< \eta <$ 3.5 &      1.646 $\pm$    0.007&       3.82 $\pm$     0.03&      11.14 $\pm$     0.13&      37.94 $\pm$     0.59&      1.372 $\pm$    0.006&       2.62 $\pm$     0.02 &      6.09 $\pm$     0.08 &     15.89 $\pm$     0.28 \\ \hline 
3.5 $< \eta <$ 4.0 &      1.621 $\pm$    0.007&       3.70 $\pm$     0.03&      10.72 $\pm$     0.13&      36.62 $\pm$     0.60&      1.332 $\pm$    0.006&       2.47 $\pm$     0.02 &      5.64 $\pm$     0.07 &     14.68 $\pm$     0.27 \\ \hline 
4.0 $< \eta <$ 4.5 &      1.591 $\pm$    0.008&       3.54 $\pm$     0.03&       9.94 $\pm$     0.14&      32.83 $\pm$     0.64&      1.284 $\pm$    0.007&       2.27 $\pm$     0.02 &      4.89 $\pm$     0.08 &     11.90 $\pm$     0.27 \\ \hline 
2.0 $< \eta <$ 4.5 &      1.676 $\pm$    0.002&       3.93 $\pm$     0.01&      11.62 $\pm$     0.07&      40.31 $\pm$     0.34&      1.585 $\pm$    0.002&       3.49 $\pm$     0.01 &      9.63 $\pm$     0.06 &     30.84 $\pm$     0.28 \\ \hline 
\hline 
 \multicolumn{9}{|c|}{WB Fit values}\\ 
 \hline-2.5 $< \eta <$ -2.0 &      1.655 $\pm$    0.009&       3.78 $\pm$     0.04&      10.49 $\pm$     0.17&      33.09 $\pm$     0.71&      1.401 $\pm$    0.008&       2.64 $\pm$     0.03 &      5.81 $\pm$     0.10 &     13.70 $\pm$     0.33 \\ \hline 
2.0 $< \eta <$ 2.5 &      1.665 $\pm$    0.006&       3.88 $\pm$     0.03&      11.22 $\pm$     0.11&      37.40 $\pm$     0.53&      1.409 $\pm$    0.005&       2.73 $\pm$     0.02 &      6.36 $\pm$     0.07 &     16.29 $\pm$     0.26 \\ \hline 
2.5 $< \eta <$ 3.0 &      1.657 $\pm$    0.006&       3.86 $\pm$     0.03&      11.17 $\pm$     0.12&      37.48 $\pm$     0.55&      1.395 $\pm$    0.006&       2.69 $\pm$     0.02 &      6.24 $\pm$     0.07 &     16.07 $\pm$     0.27 \\ \hline 
3.0 $< \eta <$ 3.5 &      1.641 $\pm$    0.006&       3.78 $\pm$     0.03&      10.94 $\pm$     0.12&      36.88 $\pm$     0.58&      1.367 $\pm$    0.006&       2.59 $\pm$     0.02 &      5.95 $\pm$     0.08 &     15.32 $\pm$     0.28 \\ \hline 
3.5 $< \eta <$ 4.0 &      1.615 $\pm$    0.007&       3.66 $\pm$     0.03&      10.45 $\pm$     0.12&      35.15 $\pm$     0.57&      1.327 $\pm$    0.006&       2.43 $\pm$     0.02 &      5.46 $\pm$     0.07 &     13.90 $\pm$     0.26 \\ \hline 
4.0 $< \eta <$ 4.5 &      1.584 $\pm$    0.008&       3.49 $\pm$     0.03&       9.62 $\pm$     0.14&      31.20 $\pm$     0.62&      1.277 $\pm$    0.007&       2.22 $\pm$     0.02 &      4.67 $\pm$     0.08 &     11.06 $\pm$     0.26 \\ \hline 
2.0 $< \eta <$ 4.5 &      1.670 $\pm$    0.002&       3.89 $\pm$     0.01&      11.44 $\pm$     0.06&      39.35 $\pm$     0.33&      1.579 $\pm$    0.002&       3.46 $\pm$     0.01 &      9.48 $\pm$     0.06 &     30.08 $\pm$     0.27 \\ \hline 
\end{tabular} 
\caption{Comparison of experimental normalized moments and normalized factorial moments of multiplicity distributions of minimum bias events with fit values from three distributions.} 
\end{table*} 

\begin{table*}[t]
\begin{tabular}{|c|c|c|c|c|c|c|c|c|} 
\hline 
$\eta$ &  $C_2$ & $C_3$  & $C_4$ & $C_5$ &  $F_2$ & $F_3$  & $F_4$ & $F_5$ \\ 
 \hline 
\hline 
 \multicolumn{9}{|c|}{Experimental values (Hard QCD)}\\ 
 \hline-2.5 $< \eta <$ -2.0 &      1.525 $\pm$    0.016&       3.01 $\pm$     0.08&       7.00 $\pm$     0.34&      18.23 $\pm$     1.36&      1.322 $\pm$    0.016&       2.17 $\pm$     0.07 &      3.99 $\pm$     0.26 &      7.78 $\pm$     0.82 \\ \hline 
2.0 $< \eta <$ 2.5 &      1.477 $\pm$    0.014&       2.80 $\pm$     0.06&       6.29 $\pm$     0.21&      15.93 $\pm$     0.78&      1.281 $\pm$    0.012&       2.01 $\pm$     0.04 &      3.58 $\pm$     0.14 &      6.86 $\pm$     0.40 \\ \hline 
2.5 $< \eta <$ 3.0 &      1.457 $\pm$    0.015&       2.72 $\pm$     0.06&       6.03 $\pm$     0.22&      15.15 $\pm$     0.79&      1.258 $\pm$    0.013&       1.93 $\pm$     0.05 &      3.38 $\pm$     0.14 &      6.39 $\pm$     0.40 \\ \hline 
3.0 $< \eta <$ 3.5 &      1.456 $\pm$    0.015&       2.73 $\pm$     0.06&       6.10 $\pm$     0.23&      15.52 $\pm$     0.85&      1.248 $\pm$    0.013&       1.91 $\pm$     0.05 &      3.34 $\pm$     0.14 &      6.38 $\pm$     0.41 \\ \hline 
3.5 $< \eta <$ 4.0 &      1.453 $\pm$    0.016&       2.72 $\pm$     0.07&       6.12 $\pm$     0.25&      15.78 $\pm$     0.94&      1.233 $\pm$    0.014&       1.86 $\pm$     0.05 &      3.23 $\pm$     0.15 &      6.18 $\pm$     0.46 \\ \hline 
4.0 $< \eta <$ 4.5 &      1.451 $\pm$    0.016&       2.72 $\pm$     0.07&       6.16 $\pm$     0.28&      16.06 $\pm$     1.11&      1.211 $\pm$    0.015&       1.79 $\pm$     0.05 &      3.07 $\pm$     0.18 &      5.84 $\pm$     0.55 \\ \hline 
2.0 $< \eta <$ 4.5 &      1.363 $\pm$    0.012&       2.33 $\pm$     0.04&       4.65 $\pm$     0.14&      10.39 $\pm$     0.44&      1.308 $\pm$    0.011&       2.10 $\pm$     0.04 &      3.92 $\pm$     0.12 &      8.04 $\pm$     0.35 \\ \hline 
\hline 
 \multicolumn{9}{|c|}{NBD Fit values}\\ 
 \hline-2.5 $< \eta <$ -2.0 &      1.535 $\pm$    0.006&       3.07 $\pm$     0.02&       7.28 $\pm$     0.09&      19.29 $\pm$     0.33&      1.337 $\pm$    0.005&       2.24 $\pm$     0.02 &      4.24 $\pm$     0.06 &      8.50 $\pm$     0.17 \\ \hline 
2.0 $< \eta <$ 2.5 &      1.479 $\pm$    0.006&       2.83 $\pm$     0.02&       6.42 $\pm$     0.07&      16.47 $\pm$     0.26&      1.285 $\pm$    0.005&       2.04 $\pm$     0.02 &      3.70 $\pm$     0.05 &      7.23 $\pm$     0.13 \\ \hline 
2.5 $< \eta <$ 3.0 &      1.459 $\pm$    0.006&       2.74 $\pm$     0.02&       6.14 $\pm$     0.07&      15.59 $\pm$     0.25&      1.261 $\pm$    0.005&       1.95 $\pm$     0.02 &      3.46 $\pm$     0.04 &      6.66 $\pm$     0.12 \\ \hline 
3.0 $< \eta <$ 3.5 &      1.456 $\pm$    0.006&       2.73 $\pm$     0.02&       6.15 $\pm$     0.08&      15.80 $\pm$     0.27&      1.248 $\pm$    0.006&       1.91 $\pm$     0.02 &      3.38 $\pm$     0.05 &      6.55 $\pm$     0.13 \\ \hline 
3.5 $< \eta <$ 4.0 &      1.451 $\pm$    0.007&       2.73 $\pm$     0.02&       6.17 $\pm$     0.08&      16.04 $\pm$     0.30&      1.231 $\pm$    0.006&       1.87 $\pm$     0.02 &      3.28 $\pm$     0.05 &      6.37 $\pm$     0.14 \\ \hline 
4.0 $< \eta <$ 4.5 &      1.450 $\pm$    0.007&       2.73 $\pm$     0.03&       6.23 $\pm$     0.09&      16.49 $\pm$     0.33&      1.211 $\pm$    0.006&       1.81 $\pm$     0.02 &      3.16 $\pm$     0.05 &      6.16 $\pm$     0.14 \\ \hline 
2.0 $< \eta <$ 4.5 &      1.365 $\pm$    0.003&       2.34 $\pm$     0.01&       4.72 $\pm$     0.03&      10.69 $\pm$     0.11&      1.309 $\pm$    0.003&       2.12 $\pm$     0.01 &      3.99 $\pm$     0.03 &      8.30 $\pm$     0.09 \\ \hline 
\hline 
 \multicolumn{9}{|c|}{SGD Fit values}\\ 
 \hline-2.5 $< \eta <$ -2.0 &      1.539 $\pm$    0.006&       3.09 $\pm$     0.02&       7.35 $\pm$     0.09&      19.53 $\pm$     0.33&      1.340 $\pm$    0.005&       2.25 $\pm$     0.02 &      4.28 $\pm$     0.06 &      8.62 $\pm$     0.16 \\ \hline 
2.0 $< \eta <$ 2.5 &      1.486 $\pm$    0.006&       2.86 $\pm$     0.02&       6.58 $\pm$     0.08&      17.07 $\pm$     0.27&      1.292 $\pm$    0.005&       2.07 $\pm$     0.02 &      3.81 $\pm$     0.05 &      7.58 $\pm$     0.13 \\ \hline 
2.5 $< \eta <$ 3.0 &      1.465 $\pm$    0.006&       2.78 $\pm$     0.02&       6.31 $\pm$     0.07&      16.30 $\pm$     0.26&      1.267 $\pm$    0.005&       1.99 $\pm$     0.02 &      3.60 $\pm$     0.04 &      7.08 $\pm$     0.12 \\ \hline 
3.0 $< \eta <$ 3.5 &      1.461 $\pm$    0.006&       2.78 $\pm$     0.02&       6.35 $\pm$     0.08&      16.66 $\pm$     0.28&      1.254 $\pm$    0.006&       1.95 $\pm$     0.02 &      3.54 $\pm$     0.05 &      7.06 $\pm$     0.13 \\ \hline 
3.5 $< \eta <$ 4.0 &      1.457 $\pm$    0.007&       2.77 $\pm$     0.03&       6.42 $\pm$     0.09&      17.15 $\pm$     0.32&      1.238 $\pm$    0.006&       1.91 $\pm$     0.02 &      3.47 $\pm$     0.05 &      7.02 $\pm$     0.14 \\ \hline 
4.0 $< \eta <$ 4.5 &      1.456 $\pm$    0.007&       2.79 $\pm$     0.03&       6.56 $\pm$     0.09&      17.99 $\pm$     0.35&      1.218 $\pm$    0.007&       1.86 $\pm$     0.02 &      3.40 $\pm$     0.05 &      7.01 $\pm$     0.15 \\ \hline 
2.0 $< \eta <$ 4.5 &      1.347 $\pm$    0.003&       2.27 $\pm$     0.01&       4.51 $\pm$     0.03&      10.09 $\pm$     0.11&      1.291 $\pm$    0.003&       2.05 $\pm$     0.01 &      3.79 $\pm$     0.03 &      7.80 $\pm$     0.09 \\ \hline 
\hline 
 \multicolumn{9}{|c|}{WB Fit values}\\ 
 \hline-2.5 $< \eta <$ -2.0 &      1.530 $\pm$    0.006&       3.04 $\pm$     0.02&       7.16 $\pm$     0.09&      18.86 $\pm$     0.34&      1.330 $\pm$    0.005&       2.21 $\pm$     0.02 &      4.13 $\pm$     0.06 &      8.22 $\pm$     0.17 \\ \hline 
2.0 $< \eta <$ 2.5 &      1.477 $\pm$    0.006&       2.80 $\pm$     0.02&       6.31 $\pm$     0.07&      15.99 $\pm$     0.26&      1.282 $\pm$    0.005&       2.02 $\pm$     0.02 &      3.60 $\pm$     0.05 &      6.90 $\pm$     0.13 \\ \hline 
2.5 $< \eta <$ 3.0 &      1.458 $\pm$    0.005&       2.72 $\pm$     0.02&       6.02 $\pm$     0.07&      15.05 $\pm$     0.24&      1.259 $\pm$    0.005&       1.93 $\pm$     0.02 &      3.35 $\pm$     0.04 &      6.30 $\pm$     0.12 \\ \hline 
3.0 $< \eta <$ 3.5 &      1.455 $\pm$    0.006&       2.71 $\pm$     0.02&       6.03 $\pm$     0.07&      15.21 $\pm$     0.26&      1.247 $\pm$    0.005&       1.89 $\pm$     0.02 &      3.28 $\pm$     0.04 &      6.17 $\pm$     0.12 \\ \hline 
3.5 $< \eta <$ 4.0 &      1.451 $\pm$    0.007&       2.71 $\pm$     0.02&       6.04 $\pm$     0.08&      15.40 $\pm$     0.29&      1.231 $\pm$    0.006&       1.84 $\pm$     0.02 &      3.17 $\pm$     0.05 &      5.96 $\pm$     0.13 \\ \hline 
4.0 $< \eta <$ 4.5 &      1.450 $\pm$    0.007&       2.71 $\pm$     0.02&       6.11 $\pm$     0.09&      15.81 $\pm$     0.32&      1.211 $\pm$    0.006&       1.79 $\pm$     0.02 &      3.04 $\pm$     0.05 &      5.71 $\pm$     0.14 \\ \hline 
2.0 $< \eta <$ 4.5 &      1.348 $\pm$    0.003&       2.24 $\pm$     0.01&       4.36 $\pm$     0.03&       9.45 $\pm$     0.11&      1.291 $\pm$    0.003&       2.02 $\pm$     0.01 &      3.64 $\pm$     0.03 &      7.22 $\pm$     0.09 \\ \hline 
\end{tabular} 
\caption{Comparison of experimental normalized moments and normalized factorial moments of multiplicity distributions of hard QCD events with fit values from three distributions.} 
\end{table*} 

\begin{table*}[t]
\begin{tabular}{|c|c|c|c|c|c|c|c|c|} 
\hline 
&  $C_2$ & $C_3$  & $C_4$ & $C_5$ &  $F_2$ & $F_3$  & $F_4$ & $F_5$ \\ 
 \hline 
\hline 
 \multicolumn{9}{|c|}{Minimum Bias}\\ 
 \hline 
Experiment &      1.665 $\pm$    0.009&       3.86 $\pm$     0.05&      11.15 $\pm$     0.25&      37.06 $\pm$     1.20&      1.579 $\pm$    0.009&       3.45 $\pm$     0.05 &      9.29 $\pm$     0.22 &     28.45 $\pm$     0.99 \\ \hline 
2NBD Fit &      1.668 $\pm$    0.003&       3.88 $\pm$     0.02&      11.22 $\pm$     0.09&      37.40 $\pm$     0.46&      1.582 $\pm$    0.003&       3.46 $\pm$     0.02 &      9.35 $\pm$     0.08 &     28.70 $\pm$     0.39 \\ \hline 
2SGD Fit  &      1.666 $\pm$    0.003&       3.88 $\pm$     0.02&      11.24 $\pm$     0.09&      37.67 $\pm$     0.42&      1.580 $\pm$    0.003&       3.46 $\pm$     0.02 &      9.37 $\pm$     0.08 &     28.94 $\pm$     0.35 \\ \hline 
2WB Fit &      1.668 $\pm$    0.003&       3.90 $\pm$     0.02&      11.43 $\pm$     0.08&      38.72 $\pm$     0.37&      1.583 $\pm$    0.003&       3.49 $\pm$     0.02 &      9.55 $\pm$     0.07 &     29.86 $\pm$     0.30 \\ \hline 
\hline 
\multicolumn{9}{|c|}{Hard QCD}\\ 
\hline 
Experiment &      1.363 $\pm$    0.012&       2.33 $\pm$     0.04&       4.65 $\pm$     0.14&      10.39 $\pm$     0.44&      1.308 $\pm$    0.011&       2.10 $\pm$     0.04 &      3.92 $\pm$     0.12 &      8.04 $\pm$     0.35 \\ \hline 
2NBD Fit &      1.364 $\pm$    0.004&       2.33 $\pm$     0.01&       4.67 $\pm$     0.05&      10.47 $\pm$     0.16&      1.308 $\pm$    0.004&       2.11 $\pm$     0.01 &      3.94 $\pm$     0.04 &      8.10 $\pm$     0.13 \\ \hline 
2SGD Fit &      1.366 $\pm$    0.004&       2.35 $\pm$     0.01&       4.74 $\pm$     0.05&      10.73 $\pm$     0.15&      1.310 $\pm$    0.004&       2.12 $\pm$     0.01 &      4.00 $\pm$     0.04 &      8.33 $\pm$     0.12 \\ \hline 
2WB Fit &      1.363 $\pm$    0.004&       2.32 $\pm$     0.01&       4.63 $\pm$     0.05&      10.27 $\pm$     0.16&      1.308 $\pm$    0.004&       2.10 $\pm$     0.01 &      3.90 $\pm$     0.04 &      7.94 $\pm$     0.13 \\ \hline 
\end{tabular} 
\caption{Comparison of experimental normalized moments and normalized factorial moments of multiplicity distributions of minimum bias and hard QCD events with fit values from 2NBD, 2SGD and 2WB distributions.} 
\end{table*} 

\begin{table*}
\begin{tabular}{|c|c|c|c|c|} 
\hline 
 \multicolumn{5}{|c|}{$<n_{ch}>$ (Minimum Bias)}\\ 
 \hline 
$\eta$ &  Data & SGD  & NBD & WB \\ 
 \hline 
-2.5 $< \eta <$ -2.0 &      3.913 $\pm$    0.070 &      3.936 $\pm$    0.019 &      3.938 $\pm$    0.019 &      3.936 $\pm$    0.019 \\ 
 \hline 
2.0 $< \eta <$ 2.5 &      3.902 $\pm$    0.042 &      3.904 $\pm$    0.012 &      3.908 $\pm$    0.012 &      3.907 $\pm$    0.012 \\ 
 \hline 
2.5 $< \eta <$ 3.0 &      3.804 $\pm$    0.042 &      3.804 $\pm$    0.012 &      3.809 $\pm$    0.012 &      3.808 $\pm$    0.012 \\ 
 \hline 
3.5 $< \eta <$ 3.5 &      3.661 $\pm$    0.041 &      3.649 $\pm$    0.012 &      3.655 $\pm$    0.012 &      3.654 $\pm$    0.012 \\ 
 \hline 
3.5 $< \eta <$ 4.0 &      3.482 $\pm$    0.040 &      3.462 $\pm$    0.012 &      3.470 $\pm$    0.012 &      3.470 $\pm$    0.012 \\ 
 \hline 
4.0 $< \eta <$ 4.5 &      3.248 $\pm$    0.038 &      3.257 $\pm$    0.014 &      3.259 $\pm$    0.014 &      3.254 $\pm$    0.014 \\ 
 \hline 
2.0 $< \eta <$ 4.5 &     11.672 $\pm$    0.099 &     11.018 $\pm$    0.021 &     11.160 $\pm$    0.019 &     11.093 $\pm$    0.020 \\ 
\hline 
 \multicolumn{5}{|c|}{$<n_{ch}>$ (Hard QCD)}\\ 
 \hline 
$\eta$ &  Data & NBD  & SGD & WB \\ 
 \hline 
-2.5 $< \eta <$ -2.0 &      4.941 $\pm$    0.095 &      5.026 $\pm$    0.023 &      5.035 $\pm$    0.024 &      4.999 $\pm$    0.025 \\ 
 \hline 
2.0 $< \eta <$ 2.5 &      5.114 $\pm$    0.069 &      5.155 $\pm$    0.020 &      5.152 $\pm$    0.020 &      5.123 $\pm$    0.021 \\ 
 \hline 
2.5 $< \eta <$ 3.0 &      5.037 $\pm$    0.069 &      5.050 $\pm$    0.019 &      5.043 $\pm$    0.020 &      5.016 $\pm$    0.020 \\ 
 \hline 
3.5 $< \eta <$ 3.5 &      4.820 $\pm$    0.067 &      4.828 $\pm$    0.019 &      4.821 $\pm$    0.020 &      4.800 $\pm$    0.020 \\ 
 \hline 
3.5 $< \eta <$ 4.0 &      4.534 $\pm$    0.065 &      4.553 $\pm$    0.019 &      4.545 $\pm$    0.020 &      4.528 $\pm$    0.020 \\ 
 \hline 
4.0 $< \eta <$ 4.5 &      4.164 $\pm$    0.064 &      4.201 $\pm$    0.018 &      4.198 $\pm$    0.019 &      4.178 $\pm$    0.020 \\ 
 \hline 
2.0 $< \eta <$ 4.5 &     17.955 $\pm$    0.235 &     17.856 $\pm$    0.051 &     17.958 $\pm$    0.052 &     17.692 $\pm$    0.061 \\ 
 \hline 

\end{tabular} 
\caption{Comparison of experimental average charged particle multiplicity $<n_{ch}>$ with the corresponding values from three fitted distributions.} 
\end{table*} 

\begin{table*}[t]
\begin{tabular}{|c|c|c|c|c|} 
\hline 
 \multicolumn{5}{|c|}{$<n_{ch}>$ (Minimum Bias)}\\ 
 \hline 
$\eta$ &  Data & 2SGD  & 2NBD & 2WB \\ 
 \hline 
2.0 $< \eta <$ 4.5 &     11.672 $\pm$    0.099 &     11.619 $\pm$    0.029 &     11.617 $\pm$    0.030 &     11.638 $\pm$    0.025 \\ 
 \hline 
 \multicolumn{5}{|c|}{$<n_{ch}>$ (Hard QCD)}\\ 
 \hline 
$\eta$ &  Data & 2SGD  & 2NBD & 2WB \\ 
 \hline 
2.0 $< \eta <$ 4.5 &     17.955 $\pm$    0.235 &     17.909 $\pm$    0.069 &     17.937 $\pm$    0.076 &     17.982 $\pm$    0.077 \\ 
\hline 
\end{tabular} 
\caption{Average charged particle multiplicity $<n_{ch}>$ for double distributions.} 
\end{table*} 


\begin{thebibliography}{50}
\bibitem{Kob}{Z. Koba, H. B. Nielsen and P. Olesen, Nucl. Phys. \textbf{B 40}, 317 (1972).}
\bibitem{NBD1}{P.K. MaeKeownand, A.W. Wolfendale, Proc. Phys. Soc. \textbf{89}, 553 (1966).}
\bibitem{NBD2}{P.Carruthers and C.C. Shih, Phys. Lett. B \textbf{127}, 242 (1983).}
\bibitem{NBD3}{M. Garetto et al., Nuovo Cimento A \textbf{38}, 38 (1977).}
\bibitem{NBD4}{A. Giovannini et al., Nuovo Cimento A \textbf{24}, 421 (1974).}
\bibitem{NBD5}{C.K. Chew, S. Dat\'{e} and D. Kiang, Mod. Phys. Lett. A \textbf{01}, No.09,  553 (1986).}
\bibitem{NBD6}{P. Carruthers, C.C. Shih, F. Zachariasen,  Int. J. Mod. Phys. A \textbf{2}, 1447 (1987).}
\bibitem{NBD7}{Michal Praszalowicz, Phys. Lett. B \textbf{704}, 566 (2011).}
\bibitem{NBD8}{M. Rybczy\'{n}ski, G. Wilk and Z. W{\l}odarczyk, Phys. Rev.D  \textbf{99}, 094045 (2019).} 
\bibitem{Aln}{G.J. Alner et al., UA5 Collaboration, Phys. Lett. B \textbf{167}, 476 (1986).}
\bibitem{Abe}{E. Abe et al., Phys. Rev. Lett. \textbf{61}, 1819 (1988).}
\bibitem{Alba}{C. Albajar et al.,UA1 Collaboration, Nuc. Phys. B \textbf{335}, 261 (1990).}
\bibitem{Fug}{C. F\"{u}glesang in Multiparticle Dynamics, Italy (1989); eds. A. Giovannini and W. Kittel, World Scientific, Singapore, 193 (1997).}
\bibitem{Gam1} {S. Hegyi, Phys. Lett. B \textbf{467}, 126 (1999).}
\bibitem{Gam2}{K. Urmossy, G.G. Barnaf\"{o}ldi and T.S. Bir\`{o}, Phys.Lett. B \textbf{701}, 111 (2011).}
\bibitem{TS1}{C. Tsallis, J. Stat.Phys. \textbf{52}, 479 (1988).}
\bibitem{TS2}{C.E. Ag\"{u}iar and T. Kodama, Physica A \textbf{320}, 271 (2003).}
\bibitem{SGD1}{Ridhi Chawla and M. Kaur, Advances in High Energy Physics \textbf{2018}, Article ID 5129341, (2018).}
\bibitem{SGD2}{Aayushi Singla and M. Kaur, Advances in High Energy Physics \textbf{2019}, Article ID 5192193, (2019).} 
\bibitem{WB1}{W. Weibull, J. Appl. Mech. \textbf{18}, 293 (1951)}.
\bibitem{WB2}{Sadhana Dash, Basanta K. Nandi, and Priyanka Sett, Phys. Rev. D \textbf{93}, 114022 (2016).}
\bibitem{CMS}{V. Khachatryan, A. M. Sirunyan et al., CMS Collab., J. High Energy Phys. \textbf{01}, 79 (2011).}
\bibitem{ATLAS}{G. Aad, B. Abbott et al., ATLAS Collaboration, New J. Phys. \textbf{13}, 053033 (2011).}
\bibitem{ALICE}{K. Aamodt, N. Abel et al., ALICE Collaboration, Eur. Phys. J. C \textbf{68}, 345 (2010).}
\bibitem{LHCb}{R. Aaij, C. Abellan Beteta et al., LHCb Collaboration, Eur. Phys. J. C \textbf{72}, 1947 (2012).}
\bibitem{SGD3}{R.~Aggarwal and M.~Kaur, Advances in High Energy Physics \textbf{2020}, Article ID 5464682, (2020).}
\bibitem{WB3}{Sadhana Dash, Basanta K. Nandi, and Priyanka Sett, Phys. Rev. D \textbf{94}, 074044 (2016).} 
\bibitem{Gio}{A. Giovannini, R. Ugoccioni, Phys. Rev. D \textbf{59,} 094020 (1999).}
\bibitem{Suz}{N. Suzuki, M. Biyajima and N. Nakajima, Phys. Rev. D \textbf{54}, 3653 (1996).}
\bibitem{IM}{I.M. Dremin, arXiv:hep-ph/0404092v, (2004).}
\bibitem{IM2}{I.M. Dremin, Phys. Lett. B \textbf{313}, 209 (1993).}
\bibitem{IM3}{I.M. Dremin, V.A. Nechitailo, JETP Lett. \textbf{58}, 881 (1993).}
\bibitem{IM4}{I.M. Dremin, Physics-Uspekhi \textbf{37}, 715 (1994); arXiv:hep-ph/9406231v1, (1994).}
\bibitem{Cap}{A. Capella and E.G. Ferreiro, Eur. Phys. J. C \textbf{72}, 1936 (2012).} 
\bibitem{SLD}{K. Abe, et al., SLD Collaboration, Phys. Lett. B \textbf{371}, 149 (1996).}
\bibitem{IM5}{I.M. Dremin, V. Arena, G. Boca et al., Phys. Lett. B \textbf{336}, 119 (1994).}
\bibitem{Gia}{G. Gianini et al.,Proc. XXIII Int. Symp. on Multiparticle Dynamics, Aspen, USA, 1993; eds M. Block and A. White: World Scientific, Singapore, 405 (1994).}
\bibitem{IM6}{I.M. Dremin, R.C. Hwa, Phys. Rev. D \textbf{49}, 5805 (1994).}
\end{thebibliography}
\end{document}